\documentclass[11pt, letterpaper, onecolumn, peerreview]{IEEEtran}

\usepackage{color}
\usepackage{colordvi}
\usepackage{graphics}
\usepackage{amsmath}
\usepackage{amssymb}
\usepackage{epsfig}
\usepackage{epic}

\newcommand{\real}{I\!\!R}

\newtheorem{theorem}{Theorem}[section]
\newtheorem{proposition}{Proposition}[section]
\newtheorem{lemma}{Lemma}[section]

\newtheorem{definition}[section]{Definition}
\newtheorem{corollary}{Corollary}[section]

\begin{document}
\title{\large \bf The necessity and sufficiency of anytime capacity for
stabilization of a linear system over a noisy communication link\\
Part II: vector systems }

\author{Anant Sahai\footnote{A.~Sahai is with the Department of
    Electrical Engineering and Computer Science, U.C.~Berkeley.},
  Sanjoy Mitter\footnote{Department of Electrical Engineering and
    Computer Science at the Massachusetts Institute of
    Technology. Support for S.K.~Mitter was provided by the Army
    Research Office under the MURI Grant: Data Fusion in Large Arrays
    of Microsensors DAAD19-00-1-0466 and the Department of Defense
    MURI Grant: Complex Adaptive Networks for Cooperative Control
    Subaward
    \#03-132 and the National Science Foundation Grant CCR-0325774.} \\
  {\small sahai@eecs.berkeley.edu, mitter@mit.edu}}


\maketitle

\begin{abstract} 
  In part I, we reviewed how Shannon's classical notion of capacity is
  not sufficient to characterize a noisy communication channel if the
  channel is intended to be used as part of a feedback loop to
  stabilize an unstable scalar linear system. While classical capacity
  is not enough, a sense of capacity (parametrized by reliability)
  called ``anytime capacity'' is both necessary and sufficient for
  channel evaluation in this context. The rate required is the log of
  the open-loop system gain and the required reliability comes from
  the desired sense of stability. Sufficiency is maintained even in
  cases with noisy observations and without any explicit feedback
  between the observer and the controller. This established the
  asymptotic equivalence between scalar stabilization problems and
  delay-universal communication problems with feedback.

  Here in part II, the vector-state generalizations are established
  and it is the magnitudes of the unstable eigenvalues that play an
  essential role. To deal with such systems, the concept of the
  anytime rate-region is introduced. This is the region of rates that
  the channel can support while still meeting potentially different
  anytime reliability targets for parallel message streams. All the
  scalar results generalize on an eigenvalue by eigenvalue basis. When
  there is no explicit feedback of the noisy channel outputs, the
  intrinsic delay of the unstable system tells us what the feedback
  delay needs to be while evaluating the anytime-rate-region for the
  channel. An example involving a binary erasure channel is used to
  illustrate how differentiated service is required in any
  separation-based control architecture.
\end{abstract}

\begin{keywords}
Real-time information theory, reliability functions, control over
noisy channels, differentiated service, feedback, anytime decoding
\end{keywords}

\IEEEpeerreviewmaketitle

\section{Introduction}

One of Shannon's key contributions was the idea that bits could be
used as a single universal currency for communication. For a vast
class of point-to-point applications, the communication aspect of the
problem can be reduced to transporting bits reliably from one point to
another where the required sense of reliability does not depend on the
application. The classical source/channel separation theorems justify
a layered communication architecture with an interface that focuses
primarily on the message rate. Rate has the advantage of being
additive in nature and so multiple applications can be supported over
a single link by simple multiplexing of the message streams. This
paradigm has been so successful in practice that researchers often
assume that it is always valid.

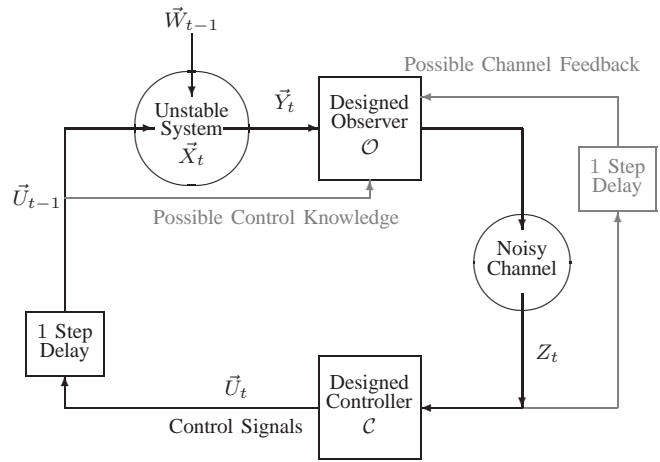
\begin{figure}[b]
\begin{center}
\setlength{\unitlength}{2100sp}%
\begingroup\makeatletter\ifx\SetFigFont\undefined%
\gdef\SetFigFont#1#2#3#4#5{%
  \reset@font\fontsize{#1}{#2pt}%
  \fontfamily{#3}\fontseries{#4}\fontshape{#5}%
  \selectfont}%
\fi\endgroup%
\begin{picture}(7643,5307)(95,-4498)
\thinlines
\put(301,-3511){\framebox(900,750){}}
\put(751,-3061){\makebox(0,0)[b]{\smash{\SetFigFont{8}{14.4}{\rmdefault}{\mddefault}{\updefault}$1$ Step}}}
\put(751,-3286){\makebox(0,0)[b]{\smash{\SetFigFont{8}{14.4}{\rmdefault}{\mddefault}{\updefault}Delay}}}
\put(6150,-2169){\oval(1142,1142)}
\put(6151,-2311){\makebox(0,0)[b]{\smash{\SetFigFont{8}{14.4}{\rmdefault}{\mddefault}{\updefault}Channel}}}
\put(6151,-2086){\makebox(0,0)[b]{\smash{\SetFigFont{8}{14.4}{\rmdefault}{\mddefault}{\updefault}Noisy}}}
\put(6451,-3311){\makebox(0,0)[b]{\smash{\SetFigFont{8}{14.4}{\rmdefault}{\mddefault}{\updefault}$Z_t$}}}
\put(3351,-361){\makebox(0,0)[b]{\smash{\SetFigFont{8}{14.4}{\rmdefault}{\mddefault}{\updefault}$\vec{Y}_t$}}}
\put(4351,-361){\makebox(0,0)[b]{\smash{\SetFigFont{8}{14.4}{\rmdefault}{\mddefault}{\updefault}Designed}}}
\put(4351,-586){\makebox(0,0)[b]{\smash{\SetFigFont{8}{14.4}{\rmdefault}{\mddefault}{\updefault}Observer}}}
\put(4351,-3661){\makebox(0,0)[b]{\smash{\SetFigFont{8}{14.4}{\rmdefault}{\mddefault}{\updefault}Designed}}}
\put(4351,-3886){\makebox(0,0)[b]{\smash{\SetFigFont{8}{14.4}{\rmdefault}{\mddefault}{\updefault}Controller}}}
\put(2251,-586){\oval(1342,1342)}
\put(3751,-1186){\framebox(1200,1200){}}
\put(3751,-4486){\framebox(1200,1200){}}
\put(2626,-586){\vector( 1, 0){1125}}
\put(2251,539){\line( 0,-1){525}}
\put(2251, 14){\vector( 0,-1){225}}
\put(4951,-586){\line( 1, 0){1200}}
\put(6151,-586){\vector( 0,-1){1200}}
\put(3751,-3886){\line(-1, 0){3000}}
\put(751,-3886){\vector( 0, 1){375}}
\put(751,-2761){\line( 0, 1){2175}}
\put(751,-586){\vector( 1, 0){1050}}
{\color[gray]{0.5}
\put(3251,-1711){\makebox(0,0)[b]{\smash{\SetFigFont{8}{14.4}{\rmdefault}{\mddefault}{\updefault}Possible Control Knowledge}}}
\put(751,-1411){\line( 1, 0){3600}}
\put(4351,-1411){\vector( 0, 1){225}}
\put(6151, 89){\makebox(0,0)[b]{\smash{\SetFigFont{8}{14.4}{\rmdefault}{\mddefault}{\updefault}Possible Channel Feedback}}}
\put(6826,-1561){\framebox(900,750){}}
\put(7276,-1111){\makebox(0,0)[b]{\smash{\SetFigFont{8}{14.4}{\rmdefault}{\mddefault}{\updefault}$1$ Step}}}
\put(7276,-1336){\makebox(0,0)[b]{\smash{\SetFigFont{8}{14.4}{\rmdefault}{\mddefault}{\updefault}Delay}}}
\put(6151,-3886){\line( 1, 0){1125}}
\put(7276,-3886){\vector( 0, 1){2325}}
\put(7276,-811){\line( 0, 1){600}}
\put(7276,-211){\vector(-1, 0){2325}}
}
\put(2776,-3736){\makebox(0,0)[b]{\smash{\SetFigFont{8}{14.4}{\rmdefault}{\mddefault}{\updefault}$\vec{U}_t$}}}
\put(2776,-4186){\makebox(0,0)[b]{\smash{\SetFigFont{8}{14.4}{\rmdefault}{\mddefault}{\updefault}Control Signals}}}
\put(4351,-886){\makebox(0,0)[b]{\smash{\SetFigFont{8}{14.4}{\rmdefault}{\mddefault}{\updefault}$\cal{O}$}}}
\put(4351,-4186){\makebox(0,0)[b]{\smash{\SetFigFont{8}{14.4}{\rmdefault}{\mddefault}{\updefault}$\cal{C}$}}}
\put(2251,-986){\makebox(0,0)[b]{\smash{\SetFigFont{8}{14.4}{\rmdefault}{\mddefault}{\updefault}$\vec{X}_t$}}}
\put(2251,614){\makebox(0,0)[b]{\smash{\SetFigFont{8}{14.4}{\rmdefault}{\mddefault}{\updefault}$\vec{W}_{t-1}$}}}
\put(426,-1486){\makebox(0,0)[b]{\smash{\SetFigFont{8}{14.4}{\rmdefault}{\mddefault}{\updefault}$\vec{U}_{t-1}$}}}
\put(2251,-451){\makebox(0,0)[b]{\smash{\SetFigFont{8}{14.4}{\rmdefault}{\mddefault}{\updefault}Unstable}}}
\put(2251,-661){\makebox(0,0)[b]{\smash{\SetFigFont{8}{14.4}{\rmdefault}{\mddefault}{\updefault}System}}}
\put(6151,-3886){\vector(-1, 0){1200}}
\put(6151,-2536){\vector( 0,-1){1350}}
\end{picture}
\end{center}
\caption{Control over a noisy communication channel. The unstable
system is persistently disturbed by $\vec{W}_t$ and must be kept
stable in closed-loop through the actions of ${\cal O, C}$.}
\label{fig:problem}
\end{figure}

Interactive applications pose a challenge to this separation based
paradigm because long delays are costly. Part I of this paper
\cite{ControlPartI} studies the requirements for the scalar version of
the interactive application illustrated in Figure~\ref{fig:problem}:
stabilization of an unstable linear system with feedback that must go
through a noisy communication channel. It turns out that message rate is
not the only relevant parameter since the underlying noisy channel
must also support enough anytime-reliability to meet the targeted
sense of stability. However, the architectural implications of this
result are unclear in the scalar case since there is only one message
stream.

To better understand the architectural requirements for interactivity
in a well defined mathematical setting, this correspondence considers
the stabilization of linear systems with a vector-valued state. Prior
work on communication-limited stabilization problems had also
considered such vector problems from a source coding
perspective. \cite{Tatikonda1} showed that the minimum rate required
is the sum of the logs of the magnitudes of the unstable eigenvalues
and \cite{NairPaper2} extends the result to certain classes of
unbounded driving disturbances. The multiparty case has begun to be
addressed in the control community under the assumption of noiseless
channels \cite{NairPaper3, YuksulBasar2}. The sequential
rate-distortion bounds calculate the best possible control performance
using a noisy channel that is perfectly matched to the unstable
open-loop system while being restricted to having a specified Shannon
capacity \cite{OurMainLQGPaper, TatikondaThesis}. Thus, the prior
necessary conditions on stabilization are only in terms of Shannon
capacity and the prior sufficient conditions required noiseless
channels.

The model of vector valued linear control systems is introduced in
Section~\ref{sec:model} and the main results are stated. Before going
into the proofs, the significance of these results is demonstrated
through an extended example in Section~\ref{sec:diffserv} involving
the stabilization of a vector-valued plant over a binary erasure channel.
For this example, stabilization is impossible unless different bits
are treated differently when it comes to transporting them across the
noisy channel. These results establish that in interactive settings, a
single ``application'' can fundamentally require different senses of
reliability for its message streams. No single number can adequately
summarize the channel and any layered architecture for reliable
communication should allow applications to individually adjust the
reliabilities on message streams. Recently, Pradhan has investigated 
block-coding reliability regions for distributed channel coding
without feedback \cite{PradhanBroadcast, PradhanMAC}. This
correspondence shows that reliability regions are interesting even in
the point-to-point case in that they are both useful and nontrivial.

Section~\ref{sec:diffserv} illustrates by example both the
implications of the results as well as how to generalize the scalar
results to vector systems with diagonalizable dynamics. The remaining
ideas involved in proving the key results are given in
Section~\ref{sec:proofsufficiency} for sufficiency and in
Section~\ref{sec:proofnecessity} for necessity. Many aspects of the
results here are straightforward generalizations from
\cite{ControlPartI} using standard linear control theory tools. To
avoid unnecessarily lengthening this correspondence, the details of
these straightforward aspects are omitted. The reader familiar with
\cite{ControlPartI, CallierDesoer} should not have any difficulty in
filling in the omitted details.

\section{The model and main results} \label{sec:model} The model is
the same as in \cite{ControlPartI} except that everything is
vector-valued. It is depicted in Figure~\ref{fig:problem}. For
convenience, all parts operate in discrete time with a common clock
for stepping through time $t$.

The $n$-dimensional state of the control system at time $t$ is
denoted $\vec{X}_t$ and evolves by
\begin{equation} \label{eqn:discretevectorsystem}
\vec{X}_{t+1} = A \vec{X}_{t} + B_u \vec{U}_{t} + B_w \vec{W}_{t}, \ \ t \geq 0.
\end{equation} 
To be interesting, the matrix $A$ should have some unstable
eigenvalues that lie strictly outside the unit circle. For the initial
condition, depending on the context we assume either a known zero
initial condition $\vec{X}_0 = \vec{0}$ or a bounded initial condition
$\|\vec{X}_t\| \leq \frac{\Omega_0}{2}$. 

Any convenient finite-dimensional norm can be used since they are all
equivalent. Consequently, this correspondence mostly assumes the
$\infty-$norm $\| \vec{X} \| = \max_i |X(i)|$ for
convenience. Throughout, subscripts are used to denote time indices,
and the $i$-th component of of the vectors is selected using $X(i)$.

The noisy channel is a probabilistic system with an input and an
output. At every time step $t$, it takes an input $a_t \in {\cal A}$
and produces an output $z_t \in {\cal Z}$ with
probability\footnote{This is a probability mass function in the case
  of discrete alphabets $\cal Z$, but is more generally an appropriate
  probability measure over the output alphabet $\cal Z$.}
$p(Z_t = z_t|a_1^t,z_1^{t-1})$ where the notation $a_1^t$ is shorthand for
the sequence $a_1, a_2, \ldots, a_t$. In general, the current channel
output is allowed to depend on all inputs so far as well as on past
outputs. 

The $m_y$ dimensional input $\vec{Y}_t$ to the observer/encoder ${\cal
  O}$ is a linear function of the state corrupted by bounded additive
noise. 
\begin{equation} \label{eqn:vectorobservation}
\vec{Y}_t = C_y \vec{X}_t + \vec{N}_t.
\end{equation}
The observer maps ${\cal O}_t: (\real^{m_y})^t \rightarrow {\cal A}$
take the observations $\vec{Y}_1^t$ and emit a channel input $a_t$. 

The channel outputs $Z_1^t$ enter the controller maps ${\cal C}_t:
{\cal Z}^t \rightarrow \real^{m_u}$ and result in the
$m_u$-dimensional control signal $\vec{U}_{t}$. 



The $\{ \vec{W}_{t} \}$ is a bounded noise/disturbance sequence taking
values in $\real^{m_w}$ s.t.~$\| \vec{W}_t \| \leq
\frac{\Omega}{2}$. Similarly, the observation noise $\{\vec{N}_t\}$ is
only assumed to be bounded so that $\| \vec{N}_t \| \leq
\frac{\Gamma}{2}$.

{\bf As in \cite{ControlPartI}, the results in this correspondence
  impose no restrictions on the individual sequences
  $\{\vec{w}_t\},\{\vec{n}_t\}$ other than remaining bounded by
  $\Omega$ and $\Gamma$.} In particular, no distribution is assumed
for these disturbances. All distributions with bounded support are
already covered by the sufficiency result here while the techniques of
\cite{OurSourceCodingPaper} can be applied to generalize the necessity
result with suitable technical conditions.

All the randomness comes from the noisy channel and any randomization
performed within the observer/encoder and controller/decoder. To be
precise, the underlying sample space $\Omega_{\mbox{sample}} =
\Omega_{\mbox{channel}} \times \Omega_{\mbox{code}}$. The channel's
randomness and the randomness available to the encoder/decoder are
assumed to be independent of each other and this is reflected in the
underlying sigma field ${\cal F}$ and probability map ${\cal
  P}$. However, once all the boxes in Figure~\ref{fig:problem} are
connected together {\em and} the individual sequences
$\{\vec{n}_t\},\{\vec{w}_i\}$ are specified, the $\{\vec{U}_t,
\vec{X}_t, \vec{Y}_t\}$ become a well defined joint random process on
the underlying probability space.

\begin{definition} \label{def:momentstable}
(Parallels Definition~2.2 in \cite{ControlPartI}) A closed-loop
dynamic system with state $\vec{X}_t$ is {\em $\eta$-stable} 
if there exists a constant $K$ s.t.~for every
$\{\vec{n}_t\},\{\vec{w}_i\}$ satisfying their bounds, the expectation
$E[\|\vec{X}_t\|^\eta] \leq K$ for all $t \geq 0$.

The expectation is taken over all the randomness in the system, 
including the noisy channel and any randomness that the controller and
observer can access. The constant $K$ may depend on the parameters of
the system including the constants $\Omega, \Gamma$, but the bound
on the $\eta$-th moment must hold for all times $t$ and uniformly over
all individual sequences for both the driving disturbance and
observation noise. 
\end{definition}
\vspace{0.1in} 

The equivalence of finite-dimensional norms guarantees that a bounded
$\eta$-moment of the $\infty-$norm of $\vec{X}$ implies a bounded
$\eta$-moment of any other norm and vice versa. This also implies that
if the $\eta$-moment is bounded in one coordinate system it is also
bounded in another coordinate system. Thus we will choose the
coordinate system best matched to the system dynamics.

The goal is to design observers ${\cal O}_t$ and controllers ${\cal
  C}_t$ that $\eta$-stabilize the system.

\subsection{Dimensionality mismatches and intrinsic delay}
Unlike the scalar case, the dimensions of
$\vec{X},\vec{U},\vec{W},\vec{Y}$ can all be different. So even
without a communication constraint, stabilizing the system in
closed-loop requires $(A,C_y)$ to be an observable pair. A pair of
matrices $(A,C)$ is observable if the matrix $[C, CA, CA^2, \ldots,
CA^{n-1}]^T$ is of full rank \cite{CallierDesoer}. This condition
assures that by combining enough raw observations, all the modes of
the linear dynamical system can be observed. The corresponding
conditions on $(A,B_u)$ and $(A, B_w)$ is that they be reachable
pairs. A pair of matrices $(A,B)$ is reachable if the matrix $[B, AB,
A^2 B, \ldots, A^{n-1}B]$ is of full rank \cite{CallierDesoer}. This
condition assures that by appropriate choice of inputs, all the modes
of the linear dynamical system can be driven to a desired state.

\begin{definition} \label{def:intrinsicdelay}
The {\em intrinsic delay} $\Theta(A,B_u,C_y)$ of a linear system is
the amount of time it takes the input to become visible at the
output. It is the minimum integer $i \geq 0$ for which $C_yA^iB_u \neq
0$. 
\end{definition}
\vspace{0.1in}
For single-input single-output (SISO) systems, this is just the
position of the first nonzero entry in the impulse response. 

%
%
%

\subsection{Anytime capacity regions}

Anytime capacity is introduced in \cite{ControlPartI} and related to
traditional channel-coding reliability functions in
\cite{OurUpperBoundPaper}. In order to state the results for systems
with vector-valued state, it is convenient to introduce the notion of
an anytime rate region. Throughout this correspondence, rates $R$ are
measured in units of bits per time step.

\begin{definition}
A rate-tuple $(R_1,R_2,\ldots)$ {\em sequential
  communication system} over a noisy channel is a channel encoder
${\cal E}$ and channel decoder ${\cal D}$ pair such that:
\begin{itemize}
\item Messages $M_{i,t}$ enter the encoder at time $t$. Message
  $M_{i,t} = S_{i,\lfloor (t-1)R_i\rfloor +1}^{\lfloor tR_i \rfloor}$
  corresponds to the $t$-th $R_i$-bit message sent in the $i$-th
  message stream and $M_{i,1}^{t-d}$ is shorthand for the sequence
  $(M_{i,1}, M_{i,2}, \ldots, M_{i,t-d})$. At the bit level, the
  $j$-th bit $S_{i,j}$ arrives at encoder $i$ at time
  $\frac{j}{R_i}$. 

\item The encoders ${\cal E}_t : {\cal Z}^{t-\theta} \times
  \{0,1\}^{\sum_{i=1}^n \lfloor R_i t \rfloor} \rightarrow {\cal A}$
  with delay-$\theta$ feedback have access to past channel outputs
  $Z_1^{t-\theta}$ in addition to the message bits, and produce a
  channel input at times $t$ based on everything it has seen so
  far.

\item The decoder ${\cal D}_t: {\cal Z}^{t} \rightarrow
  \{0,1\}^{\sum_{i=1}^n \lfloor R_i t \rfloor}$ produces updated
  estimates $\widehat{M}_{i,j}(t)$ for all $j \leq t$ based on all
  channel outputs observed till time $t$.
\end{itemize}

The {\em anytime rate region} ${\cal R}_{\mbox{any}}(\vec{\alpha})$ of
a channel is the set of rate-tuples $(R_1,R_2,\ldots)$ that the
channel can support using sequential communication. There has to exist
a uniform constant $K$ so that for each $i$, all delays $d$, all times
$t$, and {\em all possible} message sequences $\{M_{i,j}\}$,
$${\cal P}\left(\widehat{M}_{i,1}^{t-d}(t) \neq
M_{i,1}^{t-d}(t)\right) < K 2^{-\alpha_i d}.$$ 

No distribution is assumed for the bits $S_{i,j}$ and so this is
essentially a requirement on the maximum probability of error. If
common randomness is allowed, this is equivalent to assuming a uniform
distribution over the bits and an average probability of error since
the common randomness can be used to make the input look uniform by
XORing each message bit with an iid fair coin toss known to both
encoder and decoder.

The {\em $\theta$-feedback anytime rate region} refers to the rate
region when noiseless channel output feedback is available to the
encoder ${\cal E}$ with a delay of $\theta$ time units. If $\theta$ is
omitted, it is to be understood as being one.
\end{definition}
\vspace{0.1in} 

This generalizes the notion of a single feedback anytime capacity
$C_{\mbox{any}}(\alpha)$ to a rate region ${\cal
  R}_{\mbox{any}}(\vec{\alpha})$ corresponding to a vector
$\vec{\alpha}$ of anytime-reliabilities specifying how fast the
probabilities of error tend to zero with delay for the different
message streams.

Because all the message streams could simply be multiplexed
together into a single stream in which everyone has the same
reliability, it is obvious that ${\cal R}_{\mbox{any}}(\vec{\alpha})$
must contain the convex region defined by
\begin{eqnarray}
 R_i & \geq & 0, \nonumber \\
 \sum_j R_j & < & C_{\mbox{any}}(\max_j \alpha_j).
\label{eqn:trivialinnerboundregion}
\end{eqnarray}
Similarly, if $\vec{R}' \in {\cal R}_{\mbox{any}}(\vec{\alpha})$, then
any rate vector obtained by stealing rate from higher reliability
message streams and distributing it among lower reliability streams is
also going to be within the rate region.

To get a simple outer bound, just notice that any single stream could
be demultiplexed into parallel message streams and thereby achieve the
anytime reliability of at least the minimum of the parallel anytime
reliabilities. For convenience, assume that $\vec{\alpha}$ is sorted
so that $\alpha_1 \geq \alpha_2 \geq \cdots \geq \alpha_n$. This means
that ${\cal R}_{\mbox{any}}(\vec{\alpha})$ must be contained within
the region defined by the intersection of the following regions:
\begin{eqnarray}
 R_i & \geq & 0 \nonumber \\
 \sum_{j=1}^k R_j & \leq & C_{\mbox{any}}(\alpha_k)
\label{eqn:trivialouterboundregion}
\end{eqnarray}
as $i,k$ range over all the message stream indices.

\subsection{Main results}

Since the vector $\vec{\lambda}$ of unstable eigenvalues (if an
eigenvalue has multiplicity, then it should appear in $\vec{\lambda}$
multiple times) plays an important role in these results, some
shorthand notation is useful.  $\vec{\lambda}_{||}$ is used to denote
the component-wise magnitudes of the $\vec{\lambda}$.
$\log_2(\vec{\lambda_{||}})$ is used to denote the component-wise
logarithms of those magnitudes.

\begin{theorem} \label{thm:sufficiencynoisywithdelay}
For some $\vec{\epsilon} > 0$, assume a noisy finite-output-alphabet
channel such that the $(\theta+1)$-feedback anytime rate region ${\cal
R}_{\mbox{any}}(\eta \log_2 \vec{\lambda}_{||} + \vec{\epsilon})$
contains the rate vector $(\log_2(\vec{\lambda_{||}}) +
\vec{\epsilon})$. Also assume an unknown initial condition $\vec{X}_0$
that is bounded $\|\vec{X}_t\| \leq \frac{\Omega_0}{2}$. 

If the observer has access to the observations $\vec{Y}_t$ corrupted
by bounded additive noise, then any linear system with dynamics
described by (\ref{eqn:discretevectorsystem}) with unstable
eigenvalues $\vec{\lambda}$, reachable $(A,B_u)$, observable
$(A,C_y)$, intrinsic delay $\Theta(A,B_u,C_y) \leq \theta$ can be
$\eta$-stabilized by constructing an observer ${\cal O}$ and
controller ${\cal C}$ for the unstable vector system that together
achieve $E[\|\vec{X}_t\|^\eta] < K$ for {\em all sequences} of bounded
driving noise $\|\vec{W}_t\| \leq \frac{\Omega}{2}$ and {\em all
  sequences} of bounded observation noise $\|\vec{N}_t\| \leq
\frac{\Gamma}{2}$. 

If the observer maps ${\cal O}_t$ are also allowed direct access to
the past channel outputs with delay $\theta' \geq 1$, then it suffices
to just consider the $\min(\theta',\theta+1)$-feedback anytime rate
region. 
\end{theorem}
\vspace{0.1in}

By applying (\ref{eqn:trivialinnerboundregion}) to
Theorem~\ref{thm:sufficiencynoisywithdelay}, one immediately gets the
following easier to check corollary:
\begin{corollary} \label{cor:easycondition} If the sum of the
  logarithms of the magnitudes of the unstable eigenvalues of a system
  matching the conditions of
  Theorem~\ref{thm:sufficiencynoisywithdelay} is less than the
  $(\theta+1)$-delayed feedback anytime capacity $C_{\mbox{any}}(\eta
  \max_i \log_2 |\lambda_i|)$, then it is possible to $\eta$-stabilize
  the system over the noisy channel in the same sense as in
  Theorem~\ref{thm:sufficiencynoisywithdelay}.
\end{corollary}
\vspace{0.1in}

For the necessity direction, the requirements are relaxed and we only
assume that a control system exists that works with a known $\vec{0}$
initial condition and without noise in the observations since these
make the task of the observer and controller easier.
\begin{theorem} \label{thm:vectornecessity} Assume that for a given
  noisy channel, system dynamics described by
  (\ref{eqn:discretevectorsystem}) with reachable $(A,B_W)$,
  observable $(A,C_y)$, zero initial condition $\vec{X}_0 = \vec{0}$,
  eigenvalues $\vec{\lambda}$ and $\eta > 0$, that there exists an
  observer ${\cal O}$ and controller ${\cal C}$ for the unstable
  vector system that achieves $E[\|\vec{X}_t\|^\eta] < K$ for all
  sequences of bounded driving noise $\|\vec{W}_t\| \leq
  \frac{\Omega}{2}$ and all $t$.

Let $|\lambda_i| > 1$ for $i=1 \ldots l$, and let $\vec{\lambda}$ be
the $l$-dimensional vector consisting of only the exponentially
unstable eigenvalues of $A$. Then for every
$\vec{\epsilon_1},\vec{\epsilon_2} > 0$ the rate vector $(\log_2
\vec{\lambda}_{||} - \vec{\epsilon_1})$ is contained within the
$(\Theta(A,B_u,C_y)+1)$-feedback anytime rate region ${\cal
  R}_{\mbox{any}}(\eta \log_2 \vec{\lambda}_{||} - \vec{\epsilon_2})$
for this same noisy channel.
\end{theorem}
\vspace{0.1in} 

This theorem reveals that each unstable eigenvalue, no matter whether
it has its own eigenvector or not, induces a demand that the channel
be able to reliably transport a message stream. The sufficient conditions
of Theorem~\ref{thm:sufficiencynoisywithdelay} and the necessary
conditions of Theorem~\ref{thm:vectornecessity} match each other $\pm
\vec{\epsilon}$.

\section{Differentiated Service Example}\label{sec:diffserv}

This section studies a simple numeric example of a vector valued
unstable plant.  An explicit self-contained five-dimensional example
was given in \cite{SahaiAllerton00} for the binary erasure channel,
but here a simpler two-dimensional example is given that leverages the
results from \cite{OurUpperBoundPaper}. 

\begin{equation} \label{eqn:exampleA}
A = \left[ \begin{array}{cc}
2^{0.34} & 0 \\
0 & 2^{0.05}
	    \end{array}
      \right]
\end{equation}
where the observer has noiseless access to both the state $\vec{X}$
and the applied control signals $\vec{U}$. The controller can apply
any $2$-dimensional input that it wishes. Assume that the disturbance
$\vec{W}$ satisfies $\|\vec{W}_t\|_\infty \leq \frac{1}{2}$ for all
times $t$. Thus, this example consists of two independent scalar
systems that must share a single communication channel.

Section~\ref{sec:noiselessdiff} reviews how it is possible to hold
this system's state within a finite box over a noiseless channel using
total rate $R = 0.392$ consisting of one bitstream at rate $0.341$
and another bitstream of rate $0.051$. Section~\ref{sec:nodiffproblem}
considers a particular binary erasure channel and shows that if it is
used without distinguishing between the bitstreams, then third-moment
stability cannot be achieved. Section~\ref{sec:withdiffserve} shows
how a simple priority based system can distinguish between the
bitstreams and achieve third-moment stability while essentially using
the observer/controller originally designed for the noiseless link.
Finally, Section~\ref{sec:extendingexample} discusses how this
diagonal example can be transformed into a single-input single-output
control problem that suffers from the same limitations.

\subsection{Design for a noiseless channel} \label{sec:noiselessdiff}
The system defined by (\ref{eqn:exampleA}) consists of two
independent scalar systems and so each of these falls under
\cite{ControlPartI}. Since
\begin{eqnarray*}
0.341 & > & 0.34 \\
0.051 & > & 0.05,
\end{eqnarray*}
Theorem~4.1 in \cite{ControlPartI} guarantees that it is sufficient to
use two parallel bitstreams of rates $R_{1} = 0.341$ (for the
first subsystem) and $R_{2} = 0.051$
(for the second subsystem) to stabilize the system over a
noiseless channel. The total rate is $0.392$ bits per channel use. 

\subsection{Treating all bits alike} \label{sec:nodiffproblem} 
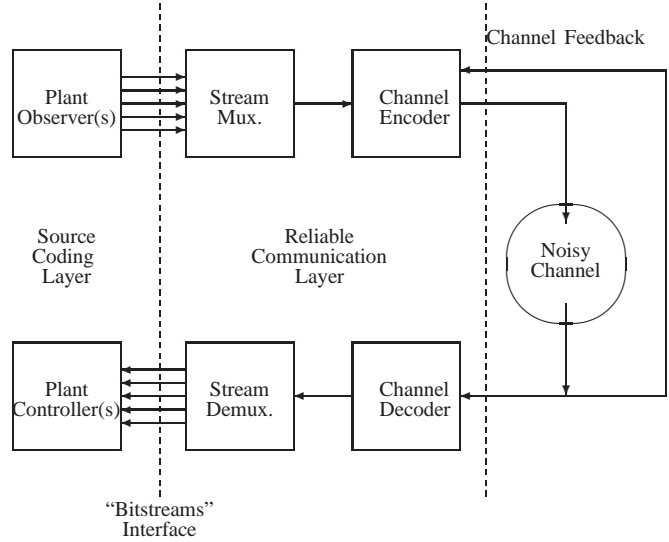
\begin{figure}
\begin{center}
\setlength{\unitlength}{2200sp}%
\begingroup\makeatletter\ifx\SetFigFont\undefined%
\gdef\SetFigFont#1#2#3#4#5{%
  \reset@font\fontsize{#1}{#2pt}%
  \fontfamily{#3}\fontseries{#4}\fontshape{#5}%
  \selectfont}%
\fi\endgroup%
\begin{picture}(7374,6012)(250,-5461)
\thinlines
\put(6450,-2394){\oval(1342,1342)}
\put(4051,-1186){\framebox(1200,1200){}}
\put(4051,-4486){\framebox(1200,1200){}}
\multiput(5551,539)(0.00000,-119.35484){47}{\line( 0,-1){ 59.677}}
\put(5251,-586){\line( 1, 0){1200}}
\put(6451,-586){\line( 0,-1){1200}}
\put(6451,-1786){\vector( 0,-1){150}}
\multiput(1876,539)(0.00000,-119.35484){47}{\line( 0,-1){ 59.677}}
\put(1426,-586){\vector( 1, 0){750}}
\put(2176,-3886){\vector(-1, 0){750}}
\put(1426,-436){\vector( 1, 0){750}}
\put(1426,-736){\vector( 1, 0){750}}
\put(1426,-886){\vector( 1, 0){750}}
\put(1426,-286){\vector( 1, 0){750}}
\put(2176,-3736){\vector(-1, 0){750}}
\put(2176,-3586){\vector(-1, 0){750}}
\put(2176,-4036){\vector(-1, 0){750}}
\put(2176,-4186){\vector(-1, 0){750}}
\put(2176,-4486){\framebox(1200,1200){}}
\put(2176,-1186){\framebox(1200,1200){}}
\put(4051,-3886){\vector(-1, 0){675}}
\put(3376,-586){\vector( 1, 0){675}}
\put(226,-4486){\framebox(1200,1200){}}
\put(226,-1186){\framebox(1200,1200){}}
\put(4351,-586){\makebox(0,0)[lb]{\smash{\SetFigFont{8}{14.4}{\rmdefault}{\mddefault}{\updefault}Channel}}}
\put(4351,-811){\makebox(0,0)[lb]{\smash{\SetFigFont{8}{14.4}{\rmdefault}{\mddefault}{\updefault}Encoder}}}
\put(4351,-3886){\makebox(0,0)[lb]{\smash{\SetFigFont{8}{14.4}{\rmdefault}{\mddefault}{\updefault}Channel}}}
\put(4351,-4111){\makebox(0,0)[lb]{\smash{\SetFigFont{8}{14.4}{\rmdefault}{\mddefault}{\updefault}Decoder}}}
\put(6451,-2311){\makebox(0,0)[b]{\smash{\SetFigFont{8}{14.4}{\rmdefault}{\mddefault}{\updefault}Noisy}}}
\put(6451,-2536){\makebox(0,0)[b]{\smash{\SetFigFont{8}{14.4}{\rmdefault}{\mddefault}{\updefault}Channel}}}
\put(2776,-586){\makebox(0,0)[b]{\smash{\SetFigFont{8}{14.4}{\rmdefault}{\mddefault}{\updefault}Stream}}}
\put(2776,-811){\makebox(0,0)[b]{\smash{\SetFigFont{8}{14.4}{\rmdefault}{\mddefault}{\updefault}Mux.}}}
\put(2776,-3886){\makebox(0,0)[b]{\smash{\SetFigFont{8}{14.4}{\rmdefault}{\mddefault}{\updefault}Stream}}}
\put(2776,-4111){\makebox(0,0)[b]{\smash{\SetFigFont{8}{14.4}{\rmdefault}{\mddefault}{\updefault}Demux.}}}
\put(826,-586){\makebox(0,0)[b]{\smash{\SetFigFont{8}{14.4}{\rmdefault}{\mddefault}{\updefault}Plant}}}
\put(826,-811){\makebox(0,0)[b]{\smash{\SetFigFont{8}{14.4}{\rmdefault}{\mddefault}{\updefault}Observer(s)}}}
\put(826,-3886){\makebox(0,0)[b]{\smash{\SetFigFont{8}{14.4}{\rmdefault}{\mddefault}{\updefault}Plant}}}
\put(826,-4111){\makebox(0,0)[b]{\smash{\SetFigFont{8}{14.4}{\rmdefault}{\mddefault}{\updefault}Controller(s)}}}
\put(1876,-5236){\makebox(0,0)[b]{\smash{\SetFigFont{8}{14.4}{\rmdefault}{\mddefault}{\updefault}``Bitstreams''}}}
\put(1876,-5461){\makebox(0,0)[b]{\smash{\SetFigFont{8}{14.4}{\rmdefault}{\mddefault}{\updefault}Interface}}}
\put(826,-2161){\makebox(0,0)[b]{\smash{\SetFigFont{8}{14.4}{\rmdefault}{\mddefault}{\updefault}Source}}}
\put(826,-2386){\makebox(0,0)[b]{\smash{\SetFigFont{8}{14.4}{\rmdefault}{\mddefault}{\updefault}Coding }}}
\put(826,-2611){\makebox(0,0)[b]{\smash{\SetFigFont{8}{14.4}{\rmdefault}{\mddefault}{\updefault}Layer}}}
\put(3676,-2161){\makebox(0,0)[b]{\smash{\SetFigFont{8}{14.4}{\rmdefault}{\mddefault}{\updefault}Reliable}}}
\put(3676,-2386){\makebox(0,0)[b]{\smash{\SetFigFont{8}{14.4}{\rmdefault}{\mddefault}{\updefault}Communication}}}
\put(3676,-2611){\makebox(0,0)[b]{\smash{\SetFigFont{8}{14.4}{\rmdefault}{\mddefault}{\updefault}Layer}}}
\put(6151,-3886){\line( 1, 0){1425}}
\put(7576,-3886){\line( 0, 1){3675}}
\put(7576,-211){\line(-1, 0){1950}}
\put(5626,-211){\vector(-1, 0){375}}
\put(6151,-3886){\vector(-1, 0){900}}
\put(6451,-2836){\vector( 0,-1){1050}}
\put(6451, 89){\makebox(0,0)[b]{\smash{\SetFigFont{8}{14.4}{\rmdefault}{\mddefault}{\updefault}Channel Feedback}}}
\end{picture}
\end{center}
\caption{Forcing all the bitstreams to get the same treatment for reliable communication}
\label{fig:multistream_mux}
\end{figure}

A
strict layering-oriented design attempts to use a virtual bit-pipe
interface to connect the observers and controllers from the previous
section. Consider a binary erasure channel with erasure probability
$\beta = 0.4$ and noiseless feedback available to the encoder. There
is clearly enough Shannon capacity since $1 - 0.4 = 0.6 >
0.392$. To minimize latency, the natural choice of coding scheme
is a single FIFO queue in which bits are retransmitted until they get
through correctly. The system is illustrated in
Figure~\ref{fig:multistream_mux}.

 
If a channel-code does not differentiate among the substreams, then
the code would have to give the same anytime reliability to all the
bits. The minimum anytime reliability required is $\alpha^* = 3 \log_2
2^{0.34} =  1.02$. For the binary erasure channel, there is exact
expression for the feedback
anytime-capacity:\cite[Theorem~3.3]{OurUpperBoundPaper} 
\begin{equation} \label{eqn:anytimeupperforerasure}
C_{\mbox{any}}(\alpha)  =  \frac{\alpha }{\alpha +
   \log_2(\frac{1-\beta}{1- \beta 2^\alpha})}.
\end{equation}
Plugging in $\beta = 0.4$ and $\alpha = 1.02$ into
(\ref{eqn:anytimeupperforerasure}) reveals that the channel can only
carry $\approx 0.38 < 0.39$ bits/channel-use with the required
reliability. Thus, it is impossible to simultaneously attain the
required rate/reliability pair by using a channel code that treats all
message bits alike.

\subsection{Differentiated service} \label{sec:withdiffserve}

\begin{figure}
\begin{center}
\setlength{\unitlength}{2200sp}%
\begingroup\makeatletter\ifx\SetFigFont\undefined%
\gdef\SetFigFont#1#2#3#4#5{%
  \reset@font\fontsize{#1}{#2pt}%
  \fontfamily{#3}\fontseries{#4}\fontshape{#5}%
  \selectfont}%
\fi\endgroup%
\begin{picture}(8128,4722)(250,-3973)
\thinlines
\put(6765,-1861){\oval(1484,1484)}
\put(5701,-511){\line( 1, 0){1050}}
\put(6751,-511){\vector( 0,-1){975}}
\put(6751,-2461){\vector( 0,-1){825}}
\put(6751,-3286){\vector(-1, 0){975}}
\put(1201, 14){\vector( 1, 0){975}}
\put(1801,164){\line( 1, 0){2700}}
\put(4501,164){\line( 0,-1){300}}
\put(4501,-136){\line(-1, 0){2700}}
\thicklines
\put(2776, 14){\line( 1, 0){1725}}
\put(1740,-2836){\line( 1, 0){661}}
\thinlines
\put(4501,-2686){\line(-1, 0){2763}}
\put(1738,-2686){\line( 0,-1){300}}
\put(1738,-2986){\line( 1, 0){2763}}
\thicklines
\put(1740,-3811){\line( 1, 0){1200}}
\thinlines
\put(4501,-3661){\line(-1, 0){2763}}
\put(1738,-3661){\line( 0,-1){300}}
\put(1738,-3961){\line( 1, 0){2763}}
\put(6751,-3286){\line( 1, 0){1050}}
\put(7801,-3286){\line( 0, 1){3225}}
\put(7801,-61){\line(-1, 0){1125}}
\put(6676,-61){\vector(-1, 0){975}}
\put(5101,164){\line( 0, 1){375}}
\put(5101,539){\line(-1, 0){900}}
\put(4201,539){\vector( 0,-1){300}}
\put(5101,-1111){\line( 0,-1){300}}
\put(5101,-1411){\line(-1, 0){900}}
\put(4201,-1411){\vector( 0, 1){300}}
\put(1651,-2836){\vector(-1, 0){450}}
\put(1651,-3811){\vector(-1, 0){450}}
\put(4876,-3961){\framebox(900,1275){}}
\put(4801,-1111){\framebox(900,1275){}}
\put(4501, 14){\vector( 1, 0){300}}
\put(4501,-961){\vector( 1, 0){300}}
\thicklines
\put(3751,-961){\line( 1, 0){750}}
\thinlines
\put(1801,-811){\line( 1, 0){2700}}
\put(4501,-811){\line( 0,-1){300}}
\put(4501,-1111){\line(-1, 0){2700}}
\put(1201,-961){\vector( 1, 0){975}}
\put(4876,-2836){\vector(-1, 0){825}}
\put(4801,-3811){\vector(-1, 0){750}}
\put(6765,-1861){\makebox(0,0)[b]{\smash{\SetFigFont{8}{14.4}{\rmdefault}{\mddefault}{\updefault}Erasure}}}
\put(6765,-2086){\makebox(0,0)[b]{\smash{\SetFigFont{8}{14.4}{\rmdefault}{\mddefault}{\updefault}Channel}}}
\put(4651,614){\makebox(0,0)[b]{\smash{\SetFigFont{8}{14.4}{\rmdefault}{\mddefault}{\updefault}Retransmission Control}}}
\put(2865,314){\makebox(0,0)[b]{\smash{\SetFigFont{8}{14.4}{\rmdefault}{\mddefault}{\updefault}Encoding Buffers}}}
\put(2851,-2536){\makebox(0,0)[b]{\smash{\SetFigFont{8}{14.4}{\rmdefault}{\mddefault}{\updefault}Decoding Buffers}}}
\put(976,-361){\makebox(0,0)[b]{\smash{\SetFigFont{8}{14.4}{\rmdefault}{\mddefault}{\updefault}Input Bitstreams}}}
\put(5326,-3211){\makebox(0,0)[b]{\smash{\SetFigFont{8}{14.4}{\rmdefault}{\mddefault}{\updefault}Priority}}}
\put(5326,-3436){\makebox(0,0)[b]{\smash{\SetFigFont{8}{14.4}{\rmdefault}{\mddefault}{\updefault}Based}}}
\put(5326,-3661){\makebox(0,0)[b]{\smash{\SetFigFont{8}{14.4}{\rmdefault}{\mddefault}{\updefault}Selection}}}
\put(5251,-361){\makebox(0,0)[b]{\smash{\SetFigFont{8}{14.4}{\rmdefault}{\mddefault}{\updefault}Priority}}}
\put(5251,-586){\makebox(0,0)[b]{\smash{\SetFigFont{8}{14.4}{\rmdefault}{\mddefault}{\updefault}Based}}}
\put(5251,-811){\makebox(0,0)[b]{\smash{\SetFigFont{8}{14.4}{\rmdefault}{\mddefault}{\updefault}Selection}}}
\put(915,-3136){\makebox(0,0)[b]{\smash{\SetFigFont{8}{14.4}{\rmdefault}{\mddefault}{\updefault}$\widehat{S}_{(i,{t-d'})}^{t-d''}$}}}
\put(915,-3436){\makebox(0,0)[b]{\smash{\SetFigFont{8}{14.4}{\rmdefault}{\mddefault}{\updefault}Variable Delay}}}
\put(915,-3661){\makebox(0,0)[b]{\smash{\SetFigFont{8}{14.4}{\rmdefault}{\mddefault}{\updefault}Output Bitstreams}}}
\put(976,-586){\makebox(0,0)[b]{\smash{\SetFigFont{8}{14.4}{\rmdefault}{\mddefault}{\updefault}$S_{(i,t)}$}}}
\end{picture}
\end{center}
\caption{The strict priority queuing strategy for discrimination
  between bitstreams. Lower priority buffers are served only if the
  higher priority ones are empty.}
\label{fig:erasure_multi}
\end{figure}
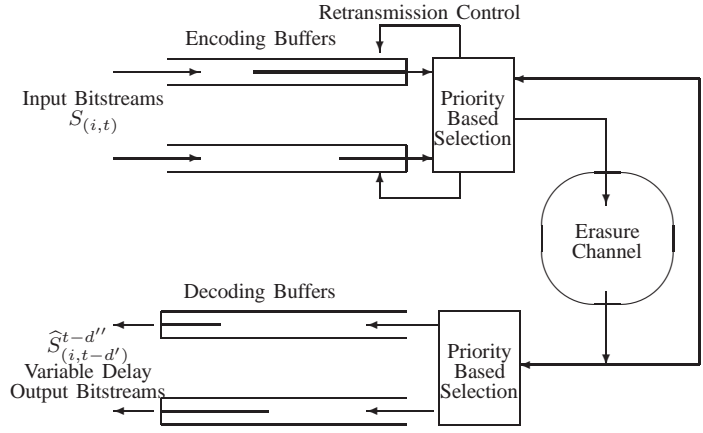

The main difficulty encountered in the previous section is that the
most challenging reliability requirement comes from the larger
eigenvalue, while the total rate requirement involves both the
eigenvalues. This section explores the idea of differentiated service
at the reliable communication layer as illustrated in the simple
priority-based scheme of Figure~\ref{fig:erasure_multi}. This is used
to give extra reliability (shorter delays) to the bitstream
corresponding to the first subsystem at the expense of lower
reliability (higher delays) for the second one.

\begin{itemize}
 \item Place bits from the different streams into
 prioritized FIFO buffers.

 \item At every channel use, transmit the oldest bit
 from the highest priority input buffer that is not empty. 

 \item If the bit is received correctly, remove it from the
 appropriate input buffer. 

 \item If there are no bits waiting in any buffer, then send a dummy
 bit across the channel.
\end{itemize}

Two priority levels are used. The higher one corresponds to the rate
$R_1 = 0.341$ bitstream coming from the first subsystem with
eigenvalue $2^{0.34}$. The lower one corresponds to the rate $R_2 =
0.051$ bitstream and corresponds to the subsystem with eigenvalue
$2^{0.05}$.

The decoder functions on a stream-by-stream basis. Since there is
noiseless feedback and the encoder's incoming bitstreams are
deterministic in their timing, the decoder can keep track of the
encoder's buffer sizes. As a result, it knows which incoming bit
belongs to which stream and can pass the received bit on to the
appropriate subsystem's controller. The sub-system controllers are
patched as in the proof of Theorem~4.1 in \cite{ControlPartI} --- they
apply $\lambda_i^d U_i(t_o)$ if their bit arrives with a delay of $d$
time-steps instead of showing up at time $t_o$ as expected. 

All that remains is to calculate a lower bound on the anytime
reliabilities delivered by such a communication scheme. 

\begin{theorem} \label{thm:lowerpriorityreliability} For the binary
  erasure channel with erasure probability $\beta > 0$ used with the
  strict two-priority encoder above, high-priority rate $0 < R_H < 1-
  \beta$ and low-priority rate $0 \leq R_L < 1 - \beta - R_H$, the
  system attains anytime-reliabilities $\alpha_H, \alpha_L$ satisfying  
\begin{eqnarray} 
\alpha_H & = & C_{\mbox{any}}^{-1}(R_H) \nonumber \\
\alpha_L & \geq & \max_{\rho \leq \rho_{HL}} E_0(\rho) - \rho R_H 
\label{eqn:strictprioritybound}
\end{eqnarray}
where
\begin{equation} \label{eqn:gallagerfunction}
E_0(\rho) = -\log_2(\beta + 2^{-\rho}(1-\beta))
\end{equation}
is the base-2 Gallager function for the BEC from \cite{gallager} and
$\rho_{HL}$ is the unique solution to
\begin{equation} \label{eqn:rhohldef}
R_H + R_L = \frac{E_0(\rho_{HL})}{\rho_{HL}}.
\end{equation}

When $R_L$ is low enough, the bound (\ref{eqn:strictprioritybound})
evaluates to the sphere-packing bound $D(1-R_H||\beta)$ for the
anytime reliability $\alpha_L$ of the lower-priority bitstream.
\end{theorem}
{\em Proof:} See Appendix~\ref{app:newprioritybound}.
\vspace{0.1in}

Numerical evaluation of the bound (\ref{eqn:strictprioritybound}) for
the rate-pair $R_H = R_1 = 0.341, R_L = R_2 = 0.051$ gives the
anytime reliabilities $\alpha_1 \approx 1.11 >
1.02$ and $\alpha_2 \geq 0.196 > 0.15$. This reveals that the both
subsystems will remain stable in the third-moment sense.

\begin{figure}[htbp]
\begin{center}
\includegraphics[width=4in,height=4.5in]{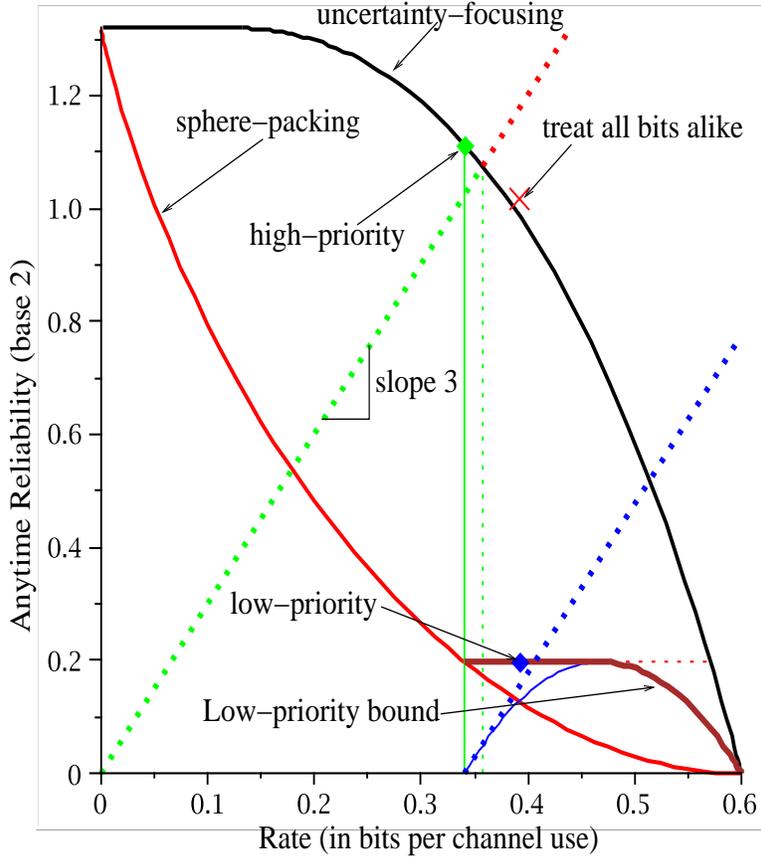}
\end{center}
\caption{The anytime capacity curve for the binary erasure channel,
  along with the sphere-packing bound and an evaluation of the
  low-priority bound for $R_H = 0.341$. The parabolic curve under the
  low-priority bound illustrates what happens as $\rho$ ranges in
  (\ref{eqn:strictprioritybound}). The maximum of the curve is
  attained at the sphere-packing bound. } 
\label{fig:erasurepriority}
\end{figure}

All of this is illustrated graphically in
Figure~\ref{fig:erasurepriority}. The diagonal lines have a slope of
$\eta=3$. The marked $\times$ on the plot has rate equal to
$0.34+0.05$ and a reliability of $3*0.34$. Since it is outside the
anytime capacity region demarcated by the BEC's uncertainty-focusing
bound, it is not achievable. The two marked $\diamond$ points
represent the reliabilities achieved by the high and low priority
streams. Notice that both are above their corresponding diagonal lines
and so the resulting closed-loop system is $3$-stable.

\subsection{Interpreting the example} \label{sec:extendingexample} The diagonal system example
given here is subject to two simple interpretations. First, it can be
interpreted as two physically distinct control systems that must share
a common bottleneck communication link. Thus, it represents an
information-theoretic example of how different interactive
applications sharing the same communication link can require
differentiated service by the reliable communication layer even in the
context of an asymptotic binary performance objective like
$\eta$-stabilization.

Alternatively, this example can be packaged into a single system with
a vector valued state. This vector-state valued system can even be at
the heart of a SISO control system. Consider the
change of coordinates matrix:
\begin{equation} \label{eqn:exampleT}
T = \left[ \begin{array}{cc}
1 & -1  \\
0 & 1      \end{array}
      \right]
\end{equation}
This transformation is used to define $\widetilde{A} = T A T^{-1}$
using (\ref{eqn:exampleT}) and (\ref{eqn:exampleA}) results in
\begin{equation} \label{eqn:exampleAtilde}
\widetilde{A} = \left[ \begin{array}{cc}
2^{0.34} & 2^{0.05}-2^{0.34}  \\
0 & 2^{0.05}      \end{array}
      \right].
\end{equation}

The $B_w$ matrix
remains the identity while
\begin{equation} \label{eqn:exampleB}
B_u = \left[ \begin{array}{c}
0 \\
1 \\
	     \end{array}
      \right]
\end{equation}
and
\begin{equation} \label{eqn:exampleC}
C_y = [1, 1].
\end{equation}
This is clearly an unstable SISO system with a scalar observation
$Y_t$ and scalar control $U_t$. As a scalar system, it has two real
poles at $2^{0.34}$ and $2^{0.05}$. Both are outside the unit circle.

It is easy to verify that $C_y$ and $C_y \widetilde{A}$ are linearly
independent and thus the system is observable. If there were neither
controls nor driving disturbances, then an observer ${\cal O}$ could
take an appropriate linear combination of two consecutive scalar
observations $Y_{i}, Y_{i+1}$ to recover the state $\vec{X}_{i}$
exactly. Explicitly, $X_{i}(1) = \frac{(2^{0.34} -
  2^{1.05})Y_i + Y_{i+1}}{2^{1.34}-2^{1.05}}$ and $X_{i}(2) =
\frac{2^{0.34} Y_i - Y_{i+1}}{2^{1.34}-2^{1.05}}$. 

The driving disturbance shows up as an additional ``noise'' in
$\vec{X}_{i+1}$. Hence the effective observation noise in $Y_{i}$ is
$N_i$ and the effective observation noise in $Y_{i+1}$ is $W_i(1) +
W_i(2) + N_{i+1}$. So the estimation error $|\widehat{X}_i(1) -
X_i(1)| \leq \frac{(2^{1.05} - 2^{0.34} + 1)\Gamma +
  2\Omega}{2(2^{1.34}-2^{1.05})}$ and similarly $|\widehat{X}_i(2) -
X_i(2)| \leq \frac{(2^{0.34} + 1)\Gamma +
  2\Omega}{2(2^{1.34}-2^{1.05})}$. This can be interpreted as a larger
$\Gamma'$ that bounds the norm of the effective observation noise. 

To support the estimation performed at the observer, the controller
could either refrain from applying any controls for two consecutive
time-instants or equivalently apply control signals that are perfectly
known to the observer. 

Similarly, the controllability conditions are satisfied since $B_u$
and $\widetilde{A} B_u$ are linearly independent. By identical
reasoning, this means that the controller can apply any desired
control to each of the underlying states by preparing scalar controls
in batches of two consecutive time units. Thus, the controller can
alternate between applying a zero control for two time units and then
applying a batched control for the next two time units.



Stabilizing the output of the SISO system clearly requires stabilizing
all of the internal states since the system is observable. Since the
internal state evolution of this system is governed by
$\widetilde{A}$, it is essentially the same as that governed by the
diagonal $A$ since the two differ only by a linear change of
coordinates. This means that differentiated service across the erasure
channel is required for this single SISO system as well!

\section{Sufficiency: Proof of
  Theorem~\ref{thm:sufficiencynoisywithdelay}} \label{sec:proofsufficiency}

Theorem~\ref{thm:sufficiencynoisywithdelay} is proven in stages. The
scalar case is in \cite{ControlPartI} and as
Section~\ref{sec:extendingexample} shows, the scalar results
immediately generalize to systems with purely diagonal dynamics
through a change of coordinates. The bounded initial condition can be
interpreted as a zero initial condition and bounded driving noise, but
for a system that starts at time $-1$. Consequently, the first new
issue concerns systems with nondiagonal Jordan blocks. After that, we
consider general systems that are reachable and observable, but where
the anytime code is assumed to be a black-box with its own access to
channel feedback.  Finally, we show how to operate a feedback anytime
code without any explicit feedback path for the channel outputs.

\subsection{Non-diagonal Jordan blocks.}
\label{sec:non-diagonal-sufficiency} 

\begin{proposition} \label{prop:jordansufficiency}
For some $\vec{\epsilon} > 0$, assume access to an anytime-code that
supports the rate vector $(\log_2(\vec{\lambda_{||}}) +
\vec{\epsilon})$ with anytime reliabilities $\eta \log_2
\vec{\lambda}_{||} + \vec{\epsilon}$. 

Consider an $n$-dimensional linear system with dynamics described by
(\ref{eqn:discretevectorsystem}) having positive real unstable
eigenvalues $\vec{\lambda}$, $A$ in block-diagonal form with each
block being upper-triangular and having a single real-valued
eigenvalue $|\lambda|$ on its diagonal, $B_u$ and $C_y$ as identity
matrices so that each state can be individually controlled, with
observations $\vec{Y}_t$ corrupted by bounded additive noise.

Then for all $\Omega \geq 0, \Gamma \geq 0$, there exists a $K > 0$ so
that the system can be $\eta$-stabilized by constructing an observer
${\cal O}$ and controller ${\cal C}$ for the unstable vector system
that together achieve $E[\|\vec{X}_t\|^\eta] < K$ for {\em all
  sequences} of bounded driving noise $\|\vec{W}_t\| \leq
\frac{\Omega}{2}$ and {\em all sequences} of bounded observation noise
$\|\vec{N}_t\| \leq \frac{\Gamma}{2}$.

Furthermore, this continues to hold even if the controller is
restricted to applying a nonzero control signal only every $n$ time
steps and the observer is similarly restricted to sample the state
every $n$ time steps.
\end{proposition} 
{\em Proof: } Because the blocks corresponding to different
eigenvalues do not interact with each other in the specified model, it
suffices to consider an $n$-dimensional square $A$ matrix that
represents a single upper-triangular block.
\begin{equation} \label{eqn:uppertriangularA}
A = \left[\begin{array}{ccccc}
            \lambda & a_{12} & a_{13} & \cdots & a_{1n} \\
            0 & \lambda & a_{23} & \cdots & a_{2n} \\
            \vdots & 0 & \ddots & \ddots &  \\
            0 & 0 & \cdots & \lambda & a_{n-1,n} \\
            0 & 0 & \cdots & 0 & \lambda \end{array}\right]
\end{equation}
      
There are two key observations. The first is that the dynamics for
the last component $X_t(n)$ are the same as in the scalar case ---
$X_{t+1}(n) = \lambda X_t(n) + W_t(n) + U_t(n)$. This faces a driving
disturbance with bound $\Omega_n = \Omega$. 

The second is that the dynamics for all the other components are given
by:
\begin{equation} \label{eqn:nonfinaljordancomponent}
X_{t+1}(i) = \lambda X_t(i) + U_t(i) + W_t(i)  + \sum_{j=i+1}^n a_{i,j} X_t(j) 
\end{equation}
Recall that the constructions in Section~IV.B of \cite{ControlPartI}
(duplicated here in Appendix~\ref{app:virtualcontrol} for reader
convenience) are based on having a {\em virtual} controlled process
$\bar{X}$ that is stabilized over a finite-rate {\em noiseless}
channel in a manner that keeps the virtual state within a
$\Delta$-sized box, no matter what the disturbances are. The observer
essentially tells the controller what controls to apply so as to do
this and protects these instructions with an anytime channel code.

Group together the weighted sum of the bounded virtual controlled
state dimensions $\sum_{j=i+1}^n a_{i,j} \bar{X}_{t}(j)$ and the net
disturbance $W_t(i))$ into a single disturbance term. The new bound on 
the disturbance is simply
$$\Omega_{i} = \Omega + \sum_{j=i+1}^n |a_{i,j}| \Delta_j$$
where $\Delta_j$ are computed recursively using the following formula
from \cite{ControlPartI}:
$$\Delta_j = \frac{\Omega_j}{1 - \lambda 2^{-R}}$$
where $R > \log_2 \lambda$. 

Since there are only a finite number of state dimensions, this shows
that ``in the box'' stabilization is possible using noiseless channels
at the appropriate rates. Just as in \cite{ControlPartI}, if used with
an anytime code, the control signal must take care to counteract the
impact of any previously erroneous control signals. Let
$\widetilde{X}_t$ represent the state at time $t$ that would result
from only the actual controls applied (no disturbances) till time
$t-1$. $A\widetilde{X}_t$ is the prediction for what that would evolve
into if a zero control were to be applied at time $t$. The current
message estimates from the anytime code at time $t$ reveal what
the desired value for $\widetilde{X}_{t+1}$ is. As $B_u$ is the
identity, the actual applied control signal is just the difference
$\widetilde{X}_{t+1} - A\widetilde{X}_t$. 

The only remaining question concerns the impact on state $j$ of such
temporary anytime decoding errors on message stream $i > j$.  Recall
that the ``impulse response'' on state $j$ of an impulse on state $i$
at time $0$ is given by $p_{i,j}(t) \lambda^t$ where $p(t)$ is some
polynomial in $t$ of order $i-j$ where the polynomial depends on the
elements of the $A$ matrix. The polynomial is bounded above by $K
(\lambda^{\epsilon})^t$ where $K$ is a constant depending on $A, n,
\lambda, \epsilon$ and $\epsilon > 0$ can be chosen as small as
desired. $K$ can further be multiplied by the (small) constant $n$ to
bound the net impact of an error on state $j$ from a decoding error in
any combination of message streams $i > j$.

Since the message streams have anytime reliability $\alpha_i$
individually, they have anytime reliability $\min_i \alpha_i$ when
considered together as a single stream. Since the relevant anytime
reliability $\min_i \alpha_i > \eta \log_2 \lambda$, we can choose
$\epsilon$ so that $\alpha_i > \eta (1 + \epsilon) \log_2 \lambda$ as
well. Thus, by the same arguments as Section~IV.D of
\cite{ControlPartI}, all the $\eta$-moments in the block will be
bounded.

Notice that $A^n$ is also upper-triangular with diagonal terms of
$\lambda^n$. Thus, by arguments identical to those of Theorem~4.4 in
\cite{ControlPartI}, the results continue to hold if both the observer
and controller are restricted to act only every $n$ time steps. \hfill
$\Box$

\subsection{Changing coordinates: complex unstable
  eigenvalues} \label{sec:complexeigensuff}

The restriction to real block-diagonal upper-triangular systems in
Proposition~\ref{prop:jordansufficiency} is easily overcome by
choosing the right coordinate frame.
\begin{proposition} \label{prop:diagonalizesuff}
Proposition~\ref{prop:jordansufficiency} holds even if the real $A$ matrix
has complex unstable eigenvalues $\vec{\lambda}$.
\end{proposition}
{\em Proof:} To avoid any complications arising from the complex
eigenvalues, the real Jordan normal form can be used
\cite{KahanJordanNotes}. This guarantees that there exists a
nonsingular real matrix $V$ so that $VAV^{-1}$ is a diagonal sum of
either traditional real-valued Jordan blocks corresponding to the real
eigenvalues and special real-valued ``rotating'' Jordan blocks
corresponding to each pair of complex-conjugate eigenvalues. The
rotating block for the pair $\lambda = \lambda_r + \lambda_j
\sqrt{-1}$ and its conjugate is a real two-by-two matrix: 
\begin{eqnarray*}
& & \left[\begin{array}{cc} \lambda_r & \lambda_j \\
                         -\lambda_j & \lambda_r 
	\end{array}\right] \\
& = &
   \left[\begin{array}{cc} |\lambda| & 0 \\
                                  0  & |\lambda| 
	 \end{array}\right]
   \left[\begin{array}{cc}
          \cos(\angle\lambda) & \sin(\angle\lambda) \\
         -\sin(\angle\lambda) & \cos(\angle\lambda) 
      \end{array}\right]
\end{eqnarray*}
which is clearly a product of a scaling matrix and a rotation matrix.
Group these two-by-two rotating blocks into a block-diagonal unitary
matrix $R$. So $VAV^{-1} = \Lambda R$ where $\Lambda$ is now a real
block-diagonal matrix whose constituent blocks are upper-triangular
and whose diagonals consist of the magnitudes of the eigenvalues.

The key is to take the rotating parts and view them through the
rotating coordinate frame that makes the system dynamics real and
block-diagonal. Transform to $\vec{X}'_{kn} = (R^{-kn} V)
\vec{X}_{kn}$ using $R^{-kn} V$ as the real time-varying coordinate
transformation.  Notice that
\begin{eqnarray*}
\vec{X}'_{k+1}
 & = & R^{-(k+1)} V \vec{X}_{k+1} \\
 & = & R^{-(k+1)} V A \vec{X}_{k} + \cdots \\
 & = & R^{-(k+1)} V A V^{-1} R^k \vec{X}'_{k} + \cdots \\
 & = & R^{-(k+1)} \Lambda R R^k \vec{X}'_{k} + \cdots \\
 & = & R^{-(k+1)} R^{k+1} \Lambda \vec{X}'_{k} + \cdots \\
 & = & \Lambda \vec{X}'_{k} + \cdots
\end{eqnarray*}
since the $\Lambda$ block-diagonal matrix commutes with the unitary
block-diagonal matrix $R$. The time-varying nature of the
transformation is due to taking powers of a unitary matrix $R$ and so
the Euclidean norm is not time-varying.

The problem in transformed coordinates falls under
Proposition~\ref{prop:jordansufficiency} and so can be
$\eta$-stabilized.  \hfill $\Box$

\subsection{Dimensionality mismatch} \label{sec:controlandobserve}
The restriction to $B_u$ and $C_y$ consisting of identity matrices so
that each state dimension can be individually controlled and observed
is also easily overcome:
\begin{proposition} \label{prop:suffcontrollableobservable}
For some $\vec{\epsilon} > 0$, assume access to an anytime-code that
supports the rate vector $(\log_2(\vec{\lambda_{||}}) +
\vec{\epsilon})$ with anytime reliabilities $\eta \log_2
\vec{\lambda}_{||} + \vec{\epsilon}$. 

Consider an $n$-dimensional linear system with dynamics described by
(\ref{eqn:discretevectorsystem}) with a real matrix $A$ with unstable
eigenvalues $\vec{\lambda}$, $(A,B_u)$ reachable, $(A,C_y)$
observable, with observations $\vec{Y}_t$ corrupted by bounded
additive noise.

Assume that the observer ${\cal O}$ has access to the applied control
signals. Then for all $\Omega \geq 0, \Gamma \geq 0$, there exists a
$K > 0$ so that the system can be $\eta$-stabilized by constructing an
observer ${\cal O}$ and controller ${\cal C}$ for the unstable vector
system that together achieve $E[\|\vec{X}_t\|^\eta] < K$ for {\em all
  sequences} of bounded driving noise $\|\vec{W}_t\| \leq
\frac{\Omega}{2}$ and {\em all sequences} of bounded observation noise
$\|\vec{N}_t\| \leq \frac{\Gamma}{2}$.
\end{proposition} 
{\em Proof:} First consider the system as though it has a $B_u$ and
$C_y$ consisting of identity matrices so that each state can be
individually controlled. Proposition~\ref{prop:diagonalizesuff} tells
us that for every $\Omega',\Gamma'$ there exists an observer ${\cal
  O}'$ and controller ${\cal C}'$ that only interact with the system
every $n$ time steps and can $\eta$-stabilize the system.

By the observability of $(A,C_y)$, it is known that there exists a
linear map $F$ so that $n$ successive measurements
$\vec{Y}_{t+1},\vec{Y}_{t+2},\ldots,\vec{Y}_{t+n}$ of the system 
suffice to recover the final state $\vec{X}_{t+n} =
F(\vec{Y}_{t+1},\vec{Y}_{t+2},\ldots,\vec{Y}_{t+n})$ if there were no
driving disturbance $\vec{W}$, observation noise $\vec{N}$ or control
signals $\vec{U}$. Since all the control signals are presumed to be
known exactly at the observer, linearity tells us that their impact on
the state $\vec{X}_{t+n}$ can be compensated for exactly. Thus, only
the effect of at most $n$ of the bounded $\vec{W}, \vec{N}$ remains.
So there exists a $\Gamma' > 0$ such that the observer has access
to a $\Gamma'$-boundedly noisy observation of the true state
$\vec{X}_{t}$ every $n$ time steps. This is used to construct an
observer ${\cal O}$ from ${\cal O}'$.

Similarly, by the controllability of $(A,B_u)$, it is known that there
exists a sequence of linear maps $G$ so that by applying controls
$\vec{U}_t = G_1(\vec{U'}),\vec{U}_{t+1} =
G_2(\vec{U'}),\ldots,\vec{U}_{t+n-1} = G_n(\vec{U'})$ in $n$
successive time-steps, the system behaves as though a single control
$\vec{U}'$ was applied to the system that had a $B_u = I$ so that all
states were immediately reachable. This is used to construct a
controller ${\cal C}$ from ${\cal C}'$.

The desired proposition follows directly. \hfill $\Box$.

\subsection{Communicating through a plant with delay}
\label{sec:communicatethrough} 

We are now in a position to prove
Theorem~\ref{thm:sufficiencynoisywithdelay} by building upon
Proposition~\ref{prop:suffcontrollableobservable}.  

{\em Proof: }Consider the assumed $(\theta+1)$-feedback anytime
code. It is clear that simply delaying the outputs of the anytime
decoder by a constant $\tau$ time-steps does not change either the
message rates or the attained anytime reliabilities. The probability
of message error merely gets worse by at most a factor of $2^{\alpha_i
  \tau}$ on stream $i$.  Set $\tau = \theta+1$.

Applying Proposition~\ref{prop:suffcontrollableobservable} gives an observer
${\cal O}'$ and controller ${\cal C}'$ satisfying the following
properties:
\begin{itemize}
 \item The closed-loop system is $\eta$-stable. 
 \item The control signal $\vec{U}_t'$ only depends on the channel
       outputs $Z_1^{t-\tau}$. 
 \item The observer ${\cal O}'$ requires access to the past channel
       outputs $Z_{t - \theta-1}$ to operate the anytime code.
 \item The observer ${\cal O}'$ requires access to the past control
       signals for its own operation. 
\end{itemize}

It suffices to give the observer access to the past channel outputs
$Z_{t - \theta-1}$ since that way, it can compute its own copy of the
control signals. If the observer has direct access to past channel
outputs, then we are done. Otherwise, the channel outputs must be
communicated back to the observer through the vector-plant using only
$\theta+1$ time steps by making the plant ``dance'' with that delay
following Section~V.B.2 in \cite{ControlPartI}. The boundedness of
both the disturbance and the observation noise means that there is a
zero-error path to communicate through the plant itself.

\begin{figure}
\begin{center}
\setlength{\unitlength}{2700sp}%
\begingroup\makeatletter\ifx\SetFigFont\undefined%
\gdef\SetFigFont#1#2#3#4#5{%
  \reset@font\fontsize{#1}{#2pt}%
  \fontfamily{#3}\fontseries{#4}\fontshape{#5}%
  \selectfont}%
\fi\endgroup%
\begin{picture}(4575,6885)(226,-6490)
\thinlines
{\color[rgb]{0,0,0}\put(1051,-2611){\vector( 0, 1){1200}}
}%
{\color[rgb]{0,0,0}\put(2401,-2161){\vector( 0, 1){750}}
}%
{\color[rgb]{0,0,0}\put(3151,164){\line( 0,-1){1275}}
\put(3151,-1111){\line(-1, 0){600}}
\put(2551,-1111){\vector( 0,-1){225}}
}%
{\color[rgb]{0,0,0}\put(3151,-1111){\line( 1, 0){600}}
\put(3751,-1111){\vector( 0,-1){225}}
}%
{\color[rgb]{0,0,0}\put(2701,-1111){\vector( 0,-1){225}}
}%
{\color[rgb]{0,0,0}\put(2851,-1111){\vector( 0,-1){225}}
}%
{\color[rgb]{0,0,0}\put(3151,-1111){\vector( 0,-1){225}}
}%
{\color[rgb]{0,0,0}\put(3301,-1111){\vector( 0,-1){225}}
}%
{\color[rgb]{0,0,0}\put(3451,-1111){\vector( 0,-1){225}}
}%
{\color[rgb]{0,0,0}\put(3601,-1111){\vector( 0,-1){225}}
}%
{\color[rgb]{0,0,0}\put(3001,-1111){\vector( 0,-1){225}}
}%
{\color[rgb]{0,0,0}\put(1801,-361){\line( 0,-1){750}}
\put(1801,-1111){\line(-1, 0){600}}
\put(1201,-1111){\vector( 0,-1){225}}
}%
{\color[rgb]{0,0,0}\put(1801,-1111){\line( 1, 0){600}}
\put(2401,-1111){\vector( 0,-1){225}}
}%
{\color[rgb]{0,0,0}\put(1351,-1111){\vector( 0,-1){225}}
}%
{\color[rgb]{0,0,0}\put(1501,-1111){\vector( 0,-1){225}}
}%
{\color[rgb]{0,0,0}\put(1801,-1111){\vector( 0,-1){225}}
}%
{\color[rgb]{0,0,0}\put(1951,-1111){\vector( 0,-1){225}}
}%
{\color[rgb]{0,0,0}\put(2101,-1111){\vector( 0,-1){225}}
}%
{\color[rgb]{0,0,0}\put(2251,-1111){\vector( 0,-1){225}}
}%
{\color[rgb]{0,0,0}\put(1651,-1111){\vector( 0,-1){225}}
}%
{\color[rgb]{0,0,0}\put(301,-4411){\vector( 1, 0){4350}}
}%
{\color[rgb]{0,0,0}\put(1501,-5611){\vector( 0, 1){1200}}
}%
{\color[rgb]{0,0,0}\put(2851,-5161){\vector( 0, 1){750}}
}%
{\color[rgb]{0,0,0}\put(301,-1411){\vector( 1, 0){4350}}
}%
{\color[rgb]{0,0,0}\put(1051,-4561){\vector( 1, 0){450}}
}%
{\color[rgb]{0,0,0}\put(2401,-4561){\vector( 1, 0){450}}
}%
{\color[rgb]{0,0,0}\put(1351,-4111){\vector( 0,-1){300}}
}%
{\color[rgb]{0,0,0}\put(1501,-4111){\vector( 0,-1){300}}
}%
{\color[rgb]{0,0,0}\put(1651,-4111){\vector( 0,-1){300}}
}%
{\color[rgb]{0,0,0}\put(1801,-4111){\vector( 0,-1){300}}
}%
{\color[rgb]{0,0,0}\put(1951,-4111){\vector( 0,-1){300}}
}%
{\color[rgb]{0,0,0}\put(2101,-4111){\vector( 0,-1){300}}
}%
{\color[rgb]{0,0,0}\put(2251,-4111){\vector( 0,-1){300}}
}%
{\color[rgb]{0,0,0}\put(2401,-4111){\vector( 0,-1){300}}
}%
{\color[rgb]{0,0,0}\put(3076,-3661){\line(-1, 0){675}}
\put(2401,-3661){\line( 0,-1){450}}
\put(2401,-4111){\line(-1, 0){1200}}
\put(1201,-4111){\vector( 0,-1){300}}
}%
\put(1051,-2836){\makebox(0,0)[b]{\smash{\SetFigFont{8}{7.2}{\rmdefault}{\mddefault}{\updefault}{\color[rgb]{0,0,0}Commit to a control sequence $\vec{U}^n$}%
}}}
\put(1051,-3061){\makebox(0,0)[b]{\smash{\SetFigFont{8}{7.2}{\rmdefault}{\mddefault}{\updefault}{\color[rgb]{0,0,0}to begin applying in the following $n$ timeslots}%
}}}
\put(1051,-3286){\makebox(0,0)[b]{\smash{\SetFigFont{8}{7.2}{\rmdefault}{\mddefault}{\updefault}{\color[rgb]{0,0,0}that stabilizes the system and counteracts}%
}}}
\put(1051,-3511){\makebox(0,0)[b]{\smash{\SetFigFont{8}{7.2}{\rmdefault}{\mddefault}{\updefault}{\color[rgb]{0,0,0}the effect of all prior modulations}%
}}}
\put(2401,-2311){\makebox(0,0)[b]{\smash{\SetFigFont{8}{7.2}{\rmdefault}{\mddefault}{\updefault}{\color[rgb]{0,0,0}Commit to next set of controls}%
}}}
\put(3151,239){\makebox(0,0)[b]{\smash{\SetFigFont{8}{7.2}{\rmdefault}{\mddefault}{\updefault}{\color[rgb]{0,0,0}Applying next set of controls}%
}}}
\put(1801,-211){\makebox(0,0)[b]{\smash{\SetFigFont{8}{7.2}{\rmdefault}{\mddefault}{\updefault}{\color[rgb]{0,0,0}Applying control $\vec{U}^n$}%
}}}
\put(1501,-6436){\makebox(0,0)[b]{\smash{\SetFigFont{8}{7.2}{\rmdefault}{\mddefault}{\updefault}{\color[rgb]{0,0,0}it is based upon}%
}}}
\put(226,-4486){\makebox(0,0)[rb]{\smash{\SetFigFont{8}{7.2}{\rmdefault}{\mddefault}{\updefault}{\color[rgb]{0,0,0}Observer timeline}%
}}}
\put(226,-1486){\makebox(0,0)[rb]{\smash{\SetFigFont{8}{7.2}{\rmdefault}{\mddefault}{\updefault}{\color[rgb]{0,0,0}Controller timeline}%
}}}
\put(1501,-5761){\makebox(0,0)[b]{\smash{\SetFigFont{8}{7.2}{\rmdefault}{\mddefault}{\updefault}{\color[rgb]{0,0,0}Knows the control sequence $\vec{U}^n$}%
}}}
\put(1276,-4700){\makebox(0,0)[b]{\smash{\SetFigFont{8}{7.2}{\rmdefault}{\mddefault}{\updefault}{\color[rgb]{0,0,0}$\Theta$}%
}}}
\put(2626,-4700){\makebox(0,0)[b]{\smash{\SetFigFont{8}{7.2}{\rmdefault}{\mddefault}{\updefault}{\color[rgb]{0,0,0}$\Theta$}%
}}}
\put(2851,-5311){\makebox(0,0)[b]{\smash{\SetFigFont{8}{7.2}{\rmdefault}{\mddefault}{\updefault}{\color[rgb]{0,0,0}Know next set of controls}%
}}}
\put(4801,-811){\makebox(0,0)[lb]{\smash{\SetFigFont{8}{7.2}{\rmdefault}{\mddefault}{\updefault}{\color[rgb]{0,0,0}While applying the controls given to}%
}}}
\put(4801,-1036){\makebox(0,0)[lb]{\smash{\SetFigFont{8}{7.2}{\rmdefault}{\mddefault}{\updefault}{\color[rgb]{0,0,0}the	left, the controller also applies}%
}}}
\put(4801,-1261){\makebox(0,0)[lb]{\smash{\SetFigFont{8}{7.2}{\rmdefault}{\mddefault}{\updefault}{\color[rgb]{0,0,0}a modulation that communicates}%
}}}
\put(4801,-1486){\makebox(0,0)[lb]{\smash{\SetFigFont{8}{7.2}{\rmdefault}{\mddefault}{\updefault}{\color[rgb]{0,0,0}the current discrete channel output back }%
}}}
\put(4801,-1711){\makebox(0,0)[lb]{\smash{\SetFigFont{8}{7.2}{\rmdefault}{\mddefault}{\updefault}{\color[rgb]{0,0,0}to the observer in $1+\Theta$ time units}%
}}}
\put(1501,-5986){\makebox(0,0)[b]{\smash{\SetFigFont{8}{7.2}{\rmdefault}{\mddefault}{\updefault}{\color[rgb]{0,0,0}the controller is applying since}%
}}}
\put(1501,-6211){\makebox(0,0)[b]{\smash{\SetFigFont{8}{7.2}{\rmdefault}{\mddefault}{\updefault}{\color[rgb]{0,0,0}it knows all the channel outputs}%
}}}
\put(3151,-3736){\makebox(0,0)[lb]{\smash{\SetFigFont{8}{7.2}{\rmdefault}{\mddefault}{\updefault}{\color[rgb]{0,0,0}and modulations to interpret }%
}}}
\put(3151,-3511){\makebox(0,0)[lb]{\smash{\SetFigFont{8}{7.2}{\rmdefault}{\mddefault}{\updefault}{\color[rgb]{0,0,0}Use knowledge of controls $\vec{U}^n$}%
}}}
\put(3151,-3961){\makebox(0,0)[lb]{\smash{\SetFigFont{8}{7.2}{\rmdefault}{\mddefault}{\updefault}{\color[rgb]{0,0,0}observations $\vec{Y}^n$ to estimate $\vec{X}$}%
}}}
\end{picture}
\end{center}
\caption{When viewing time in blocks of $n$, the controller is
required to commit to its primary controls $1$ time step before
actually putting them into effect. This way, by the time the observer
can first see the effect of these controls, it already knows exactly
what that effect is going to be since it knows all the channel outputs
that the controls were based upon.} \label{fig:communicating_back}
\end{figure}
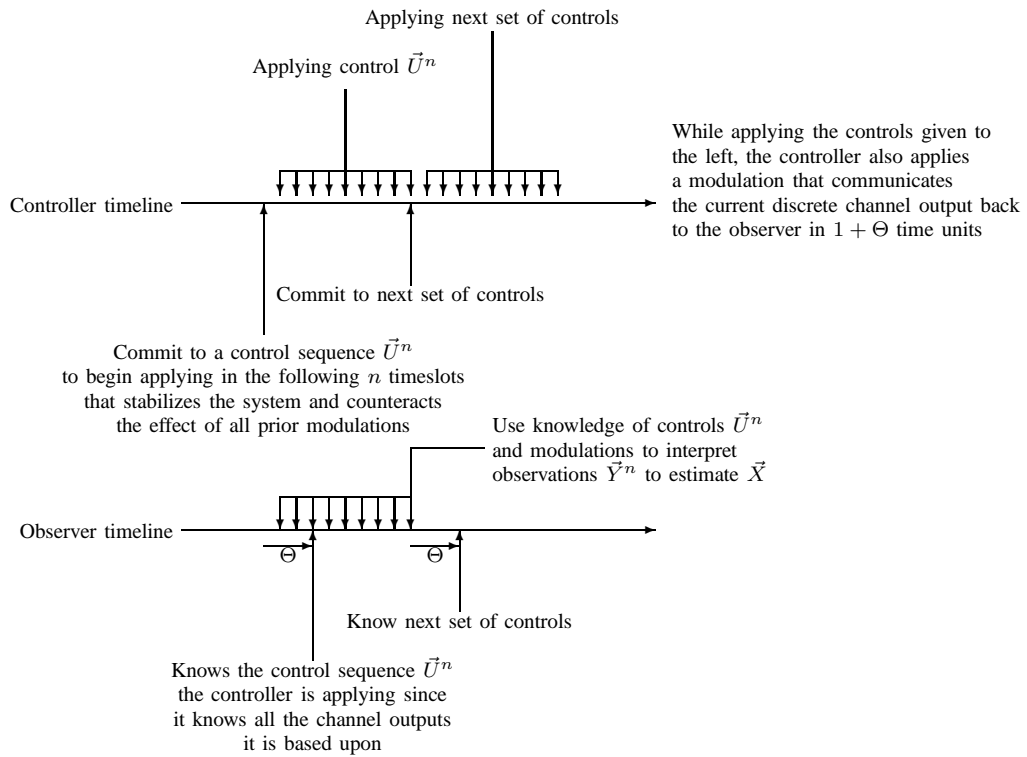

The key idea is illustrated in
Figure~\ref{fig:communicating_back}. Without loss of generality,
assume that $C_yA^{\Theta}B_u$ has a nonzero element in its first
column. Let $\vec{\psi}$ be the response of the system at time
$\Theta$ (the intrinsic delay through the plant) when fed an input of
the $(1,0,\ldots,0)^T$ vector at time $1$. Let $\bar{\psi}$ be the
maximum of $|\psi_1|,|\psi_2|,\ldots,|\psi_m|$. Let $\Gamma''$ be the
maximum magnitude of the effective observation noise at the receiver
after accounting for the combined bounded uncertainties in both the
true observation noise as well as the driving disturbances. 

Associate the finite channel output alphabet ${\cal Z}$ with the
positive integers $1,2,\ldots,|{\cal Z}|$.  Add
$3\frac{\Gamma''}{\bar{\psi}} Z_t$ to the first dimension of the
control signal $\vec{U}'_t$ before applying it. In $\Theta$ time
steps, the response will show up at the observer as a shift in
$\vec{Y}_{t+\Theta+1}$ that is unmistakably decodable to recover
$Z_{t-\Theta-1}$ exactly.

As in Section~V.B.2 of \cite{ControlPartI}, the controller uses the
controllability of $(A,B_u)$ to superimpose another control input (in
blocks of $n$ time-steps) whose purpose is to prevent the past
communication-oriented controls $3\frac{\Gamma''}{\bar{\psi}} Z_t$
from continuing to propagate unstably through the system
dynamics. Since this is only a function of past channel channel
outputs, its effect can also be removed from the observations at the
observer.

Since the additional communication-oriented control signals only have
an impact that lasts for at most $2n$ time steps, the $\eta$-stability
of the closed-loop system is unchanged and the theorem is
proved. \hfill $\Box$

\section{Necessity: Proof of Theorem~\ref{thm:vectornecessity}}
\label{sec:proofnecessity} The extension of the scalar-case
Theorem~3.3 in \cite{ControlPartI} to the vector case is largely
straightforward and for the most part, the same arguments that worked
in Section~\ref{sec:proofsufficiency} apply on the necessity side ---
with the controllability of $(A,B_w)$ playing the same role here that
controllability of $(A,B_u)$ did in
Section~\ref{sec:proofsufficiency}. Observability is not an issue
since the goal is to simulate an unstable system driven by bounded
disturbances by using the message bits as well as the $\Theta+1$
delayed channel outputs. The embedding is such that the uncontrolled
process (without the $\vec{U}$ controls) grows exponentially with time
and has high-order bits representing message bits from a long time
ago. Since the controlled process is the sum of the uncontrolled
process and the undisturbed process (without the $\vec{W}$
disturbances), the size of $\|\vec{X}_t\|$ captures the extent to
which the controller knows the embedded message bits.

Controllability of $(A,B_w)$ can be used to apply any desired input
sequence to the individual eigenstates, at the expense of a smaller
bound $\Omega$ since the original disturbance constraint might turn
into something smaller after passing through the linear mapping
induced by the reachability Grammian $[B, AB_w, A^2 B_w, \ldots,
A^{n-1}B_w]$. This leaves only two non-obvious issues:
\begin{itemize}
 \item Dealing with channel feedback that is delayed by $\Theta+1$ 
       time-steps. 
 \item Dealing with non-diagonal Jordan blocks.
\end{itemize}

Otherwise, the problem reduces by a change of coordinates to parallel
scalar systems and Theorem~3.3 in \cite{ControlPartI} gives the
desired result. To avoid repeating the same arguments as the previous
section and \cite{ControlPartI}, we focus here only on the new issues.

\subsection{Using delayed feedback to simulate the plant}
The key idea is that we do not need to feed the simulated plant state
$\vec{X}$ to the observer ${\cal O}$, just the simulated plant
observation. In order to generate the simulated $\vec{Y}_{t+1}$, the
exact $\vec{U}$ values are only needed through time
$t-\Theta(A,B_u,C_y)$ since controls after that point have not become
visible yet at the plant output. In running the simulated control
system at the anytime encoder, a delay of $1+\Theta(A,B_u,C_y)$ can
therefore be tolerated rather than the unit delay assumed while
proving Theorem~3.3 in \cite{ControlPartI}.

\subsection{Non-diagonal Jordan blocks}
It suffices to consider a single real upper-triangular block
(\ref{eqn:uppertriangularA}) since the real Jordan form decouples a
general vector problem into such components by a rotating change of
coordinates. 

The $n$ parallel bitstreams are encoded independently at rates
$\log_2 \lambda > R = R_1 = R_2 = \cdots = R_n $ into the
simulated individual driving disturbances $W_i(t)$ using the simulator
given by equation (6) in the proof of Theorem~3.3 in
\cite{ControlPartI}. The new challenge arises at the decoder.

Notice that the last state $X_t(n)$ is just like the scalar case and
only depends on its own bitstream. However, all the other states have
a mixture of bitstreams inside of them since the later states enter as
interfering inputs into the earlier states. As a result, the decoding
algorithm given in Section~III.B.2 of \cite{ControlPartI} will not
work on those other states without modification.

The decoding strategy in the upper-triangular case changes to be
successive-decoding in the style of decoding for the stronger user in
a degraded broadcast channel \cite{CoverThomas}. Explicitly, the
decoding procedure is as follows for every given time $t$ at the
decoder:
\begin{enumerate} 
  \item Set $i=n$. Set $D_t(j) = -\widetilde{X}_t(j)$ for all $j$ where
  $\widetilde{X}(j)$ represents the $j$-th component of the system in
  transformed coordinates driven only by the control inputs
  $\vec{U}'$, not the disturbances $\vec{W}'$. This is what is
  available at the decoder.
  \item Decode the bits on the $i$th stream using the algorithm of
  Section~III.B.2 of \cite{ControlPartI} applied to $D_t(i)$.
  \item Subtract the impact of these decoded bits from $D_t(k)$ for
  every $k < i$. 
  \item Decrement $i$ and goto step 2.
\end{enumerate}

Notice that if all the bits decoded upto a point are correct, then
when decoding the bits on the $i$th stream (using $D_t(k)$ as the
input to the bit-extraction algorithm of Section~III.B.2 of
\cite{ControlPartI}), the $D_t(k)$ will contain exactly what it would
have contained had the $A$ matrix been diagonal. Consequently, the
error probability calculations done in \cite{ControlPartI} would
apply. However, this successive decoding strategy has the possibility
of propagating errors between streams and so the error propagation
must be accounted for.

The goal of the $\vec{\epsilon_2}$ is to allow a slightly lower sense
of reliability for the early streams within a block. Equation (11) in
\cite{ControlPartI} tells how much of a deviation in $D_t(i)$ can be
tolerated without an error in decoding bits from before $d$ time steps
ago. Repeated here:
$$ \mbox{gap}_t(i) =  \inf_{\bar{S} : \bar{S}_i \neq S_i} |\check{X}_t(S) -
 \check{X}_t(\bar{S})| > 
 \left\{\begin{array}{ll} \lambda^{t-\frac{i}{R}} 
 \left(\frac{2 \gamma  \epsilon_1}{1+\epsilon_1}\right) & \mbox{if }i
 \leq \lfloor Rt \rfloor \\
 0 & \mbox{otherwise}	  
        \end{array}\right.
$$
where $\gamma = \frac{\Omega}{2 \lambda^{1 + \frac{1}{R}}}, \epsilon_1
= 2^{\frac{\log_2 \lambda}{R}} - 2$ are constants defined in
\cite{ControlPartI} that depend on the message rate $R$ and the size
$\Omega$ allowed while simulating the driving disturbances $W$.
 
To get an upper bound on the probability of error, allocate half of
that maximum deviation $\mbox{gap}_t(i)$ into $n-i+1$ equally-sized
pieces. So each  allocated margin is at most of size $\frac{\gamma
  \epsilon_1}  {(n-i+1)(1+\epsilon_1)} 2^{d \log_2 \lambda}$ when
considering a bit with delay $d$. The first $n - i$ of them correspond
to allowances for error propagation from later streams. The final
piece corresponds to what is allowed from the controlled state at this
level. For purposes of bounding, an error is declared whenever any one
of these pieces exceeds its allocation.

The following Lemma shows that error
propagation can cause a total deviation only a little larger than
$\lambda^d$ on an exponential scale.

\begin{lemma} \label{lem:other_stream_error_lemma} Consider a real
  Jordan block corresponding to $\lambda$ and time $t$. Suppose that
  there are only decoding errors in a stream $i > j$ occurring for
  bits corresponding to times after $t-d$ and there are no decoding
  errors on bits whose delays exceed $d$.

  Then for every $\epsilon' > 0$, there exists a $K' > 0$ so that the
  maximum magnitude deviation of $D_j$ due to the decoding errors in
  stream $i$ is bounded by $K' 2^{d (1 + \epsilon')\log_2 \lambda} =
  K' \lambda^{(1+\epsilon')d}$.
\end{lemma}
{\em Proof:} See Appendix~\ref{app:lemma}.
\vspace{0.1in}

Using Lemma~\ref{lem:other_stream_error_lemma} and setting $d'$ to the
delay corresponding to the first bit-error in the other stream, the
allocated margin can be set equal to the propagation allowance:
\begin{eqnarray*}
\frac{\gamma \epsilon_1}{(n-i+1)(1+\epsilon_1)} 
2^{d \log_2 \lambda} & = & 
K' 2^{d' (1 + \epsilon')\log_2 \lambda} \\
\frac{\gamma \epsilon_1}
{K'(n-i+1)(1+\epsilon_1)} 2^{d \log_2 \lambda} & = & 
2^{d' (1 + \epsilon')\log_2 \lambda} \\
\frac{\log_2(\frac{\gamma \epsilon_1} {K'(n-i+1)(1+\epsilon_1)})}{(1 +
\epsilon')\log_2 \lambda}  + d \frac{1}{1 + \epsilon'} & = & d' \\
K''  + d \frac{1}{1 + \epsilon'} & = & d'
\end{eqnarray*}
The key point to notice is that the tolerated delay $d'$ on the other
streams is a constant $K''$ plus a term that is almost equal to $d$.

Consequently, the probability of error on stream $i$ for bits at delay
$d$ or more is upper-bounded by
$${\cal P}\left(|X_t(i)| \geq \frac{\gamma \epsilon_1}
{(n-i+1)(1+\epsilon_1)} 2^{d \log_2 \lambda}\right) + \sum_{j=i+1}^n {\cal P}(\mbox{Stream }j\mbox{ has
an error at position }K''  + d \frac{1}{1 + \epsilon'}\mbox{ or
earlier})$$

Finite induction completes the proof. The base case, $i=n$ is obvious
since it is just the scalar case by itself. Now assume that for every
$j>i$,
$${\cal P}(\mbox{Stream }j\mbox{ has
an error at position }d\mbox{ or earlier}) \leq K'''_j 2^{-d
\frac{\eta \log_2 \lambda}{(1+\epsilon')^{n-j}}}$$
With the induction hypothesis and base case in hand, consider $i$ and
use Markov's inequality since the $\eta$-moment is bounded:
\begin{eqnarray*} 
& & {\cal P}(\mbox{Stream }i\mbox{ has an error at position }d\mbox{ or
    earlier}) \\
& \leq & 
{\cal P}(|X_t(i)| \geq \frac{\gamma
    \epsilon_1} {(n-i+1)(1+\epsilon_1)} 2^{d \log_2 \lambda}) + 
\sum_{j=i+1}^n K'''_j 2^{-(K'' + d\frac{1}{1 + \epsilon'})\frac{\eta \log_2
    \lambda}{(1+\epsilon')^{n-j}}} \\
& = & 
{\cal P}(|X_t(i)| \geq \frac{\gamma
    \epsilon_1} {(n-i+1)(1+\epsilon_1)} 2^{d \log_2 \lambda}) + 
\sum_{j=i+1}^n K'''_j 
2^{-K''\frac{\eta \log_2 \lambda}{(1+\epsilon')^{n-j}}}
2^{-d\frac{\eta \log_2 \lambda}{(1+\epsilon')^{n-j+1}}} \\
& \leq & 
K'''' 2^{-d \eta \log_2 \lambda} + 
K'''''_{i} 2^{-d\frac{\eta \log_2 \lambda}{(1+\epsilon')^{n-i}}} \\
& \leq & 
K''''''_{i} 2^{-d\frac{\eta \log_2 \lambda}{(1+\epsilon')^{n-i}}} \\
\end{eqnarray*}
where we used the induction hypothesis, the proof of Theorem~3.3 in
\cite{ControlPartI} and the fact that a finite sum of exponentials is
bounded by a constant times the slowest exponential. Since $\epsilon'$
was arbitrary and $n$ is finite, this proves the theorem since we can
get as close as we want to $\alpha = \eta \log_2 \lambda$ in anytime
reliability. \hfill $\Box$ \vspace{0.15in}

\section{Conclusions}
Theorems \ref{thm:sufficiencynoisywithdelay} and
\ref{thm:vectornecessity} reveal that the problem of stabilizing a
linear vector plant over a noisy channel is intimately connected to
the problem of reliable anytime communication of parallel message
streams over a noisy channel with feedback. The anytime-capacity
region of a channel with feedback is the key to understanding whether
or not it is possible to stabilize an unstable linear system over that
noisy channel. The two problems are related through three
parameters. The primary role is played by the magnitudes of the
unstable eigenvalues since their logs determine the required
rates. The target moment $\eta$ multiplies these logs to give the
required anytime reliabilities. Finally, the intrinsic delay
$\Theta(A,B_u,C_y)$ tells us the noiseless feedback delay to use while
evaluating the required anytime reliabilities when explicit channel
feedback is not available and all feedback must be implicitly through
the system itself.

To stabilize a system, it is sometimes necessary to treat some bits as
being more time-sensitive than others. Though the example in
Section~\ref{sec:diffserv} was crafted with the binary erasure channel
in mind, we believe that similar examples should exist for most
channels. However, there are also special channels for which such
examples do not exist. In particular, the average-power constrained
AWGN channel with noiseless feedback is special. As shown in
\cite{ControlPartI}, the AWGN channel has a feedback anytime capacity
equal to its Shannon capacity regardless of $\alpha$. The need for
differentiated service can only exist when there is a nontrivial
tradeoff between rate and reliability. 

Despite this, the ideas of this correspondence are significant even in
the case of AWGN channels. They show that stabilization (of all
moments) is possible over an adequate capacity AWGN channel with
noiseless feedback even when there is a dimensionality mismatch
between the channel and the plant. Prior results involving only linear
control theoretic techniques could not reach the capacity bound for
cases in which the dimension of the unstable plant was different than
the dimension of the channel \cite{OurMainLQGPaper}.

It should also be immediately clear that all the arguments given in
\cite{ControlPartI} on continuous-time models also apply in the
context of vector-valued states. Standard results on sampling linear
systems tell us that in the continuous-time case, the role of the
magnitude of the unstable eigenvalues is played by the positive real
part of the unstable eigenvalues. Similarly, all the results regarding
the almost-sure sense of stabilization when there is no persistent
disturbance also carry over directly with no differentiated service
required among the unstable eigenvalues. In addition, it is easy to
extend the suboptimal but ``nearly memoryless'' simple random observer
strategy of Theorem~5.2 of \cite{ControlPartI} to the vector context
by randomly labeling a lattice-based quantization of $n$ successive
observations $\vec{Y}_t$. This is suboptimal because it treats all
dimensions alike and also does not take advantage of the feedback to
improve the anytime reliability of the channel.

It should be noted that because the results given here apply for
general state-space models, they also apply to all equivalent linear
models. In particular, they apply to the case of control systems
modeled using ARMA models or with rational open-loop transfer
functions of any finite order. Assuming that there is no pole/zero
cancellation, such results can be obtained using standard linear
techniques establishing the equivalence of SISO models to the general
state-space forms considered here. In those cases, the unstable
eigenvalues of the state-space model correspond to the unstable poles
(together with their multiplicities) of the ARMA model. The intrinsic
delay corresponds to the number of leading zeros in the impulse
response, i.e.~the multiplicity of the zero at $z=\infty$.

The primary limitation of the results so far is that they only cover
the binary question of whether the plant is stabilizable in the
$\eta-$moment sense or not. They do not address the issue of
performance. In \cite{OurSourceCodingPaper}, we have a clean approach
to performance for the related scalar estimation problem using
rate-distortion techniques. The linear systems techniques of this
correspondence apply directly to the estimation problem there and can
generalize those results naturally to the vector case. In particular,
it is straightforward, but somewhat cumbersome, to apply these
techniques to completely solve all the nonstationary auto-regressive
cases left open in \cite{Gray70}.

For the estimation problem of \cite{OurSourceCodingPaper} where the
limit of large estimation delays does not inherently degrade
performance, it turns out that $l$ parallel bitstreams corresponding
to each unstable eigenvalue are required, each of rate $> \log_2
|\lambda_i|$, together with one residual bitstream that is used to
boost performance in the end-to-end distortion sense. The unstable
streams all require anytime reliability in the sense of
Theorem~\ref{thm:vectornecessity} while the residual stream just
requires Shannon's traditional reliability. Since there are no control
signals in the case of estimation, intrinsic delay plays no role
there.

A second limitation of the results so far is that there are no good
inner or outer bounds on the anytime rate and reliability regions
beyond the ones for the single-rate/reliability region
\cite{OurUpperBoundPaper}. However, even without such bounds, we have
learned something nontrivial about the relative difficulty of
different stabilization problems. For example, consider a scalar
system with a single unstable eigenvalue of $\lambda = 8$ as compared
to a vector system with three unstable eigenvalues, all of which
are $\lambda_i = 2$. From a total rate perspective, the two appear
identical requiring at least $3$ bits per unit time. However, they can
be distinguished based on the anytime-reliability they require. The
scalar case requires anytime-reliability $\alpha > 3\eta$ while the
vector case can make do with any $\alpha > \eta$. Since the three
eigenvalues are identical in the vector case, there is also no need to
prioritize any one of them over the others and thus we can interpret
the ``vector-advantage'' as being a factor reduction in the
anytime-reliability required. Thus, in the precise sense of
Section~VII of \cite{ControlPartI}, vector-stabilization problems are
easier than the scalar-stabilization problem having the same rate
requirement.\footnote{This vector advantage in terms of required
  anytime reliability is even more surprising in light of the
  performance bounds in terms of rate only. \cite{OurMainLQGPaper}
  gives explicit bounds on the squared-error performance using
  sequential distortion-rate theory. Suppose the $\lambda=8$ scalar
  plant was driven by a standard iid Gaussian disturbance while the
  vector plant was diagonal and driven by three iid Gaussians each of
  variance $\frac{1}{3}$. For a given rate $R$ (in bits), the
  sequential distortion-rate bound on $E[|X_t|^2]$ is
  $\frac{1}{1-4^{3-R}}$ for the scalar system while it is
  $\frac{1}{1-4^{1-\frac{R}{3}}}$ for the vector system. For a given
  rate, the second-moment performance of the vector system is {\em
    worse} than the scalar one. For example, at rate $4$ the scalar
  one gets to $\approx 1.33$ while the vector one is $\approx
  2.70$. At high rates, the two approach each other in terms of
  second-moment performance but the anytime-reliability requirements
  for the scalar system remain much higher.} It seems that spreading
the potential growth of the process across many independent dimensions
reduces the reliability requirements demanded from the noisy channel.

\appendices 

\section{Bounding the anytime reliability region of the strict
  priority queue} \label{app:newprioritybound}

The proof of Theorem~\ref{thm:lowerpriorityreliability} builds upon
the proof of Theorem~3.3 in \cite{OurUpperBoundPaper}. There, the
anytime capacity of the binary erasure channel with noiseless
instantaneous feedback is computed and shown to achieve the
uncertainty-focusing bound which is given parametrically by
\begin{equation} \label{eqn:parametricfocusing}
\alpha = E_0(\rho) ~~,~~ R = \frac{E_0(\rho)}{\rho}.
\end{equation}
Furthermore, it is shown in \cite{OurUpperBoundPaper} that this
reliability is attained by the strategy of placing the bits as they
deterministically arrive into a FIFO queue that is drained by $1$ bit
every time the BEC is successful.

In such a code, there is a one-to-one mapping between the queue-length
distribution and the delay distribution. Ignoring integer effects for
the sake of notational convenience, the event that bit $R(t-d)$
experiences a delay of larger than $d$ is equivalent to the event that
the queue contains at least $Rd$ bits at time $t$. Since the marginal
for delay has an exponential tail governed by the exponent $\alpha$,
this means that the steady-state queue-length $Q$ has a tail governed
by $\frac{\alpha}{R}$. Mathematically, $\forall \epsilon > 0, \exists
K > 0$ so
\begin{equation} \label{eqn:queuetail1}
{\cal P}(Q \geq q) \leq K 2^{-(\frac{\alpha(R)}{R} - \epsilon) q}
\end{equation}
where $\alpha(R)$ is the delay-reliability attained at rate $R$ as
governed parametrically by (\ref{eqn:parametricfocusing}). Defining
$\rho_R$ as the unique $\rho$ that satisfies
(\ref{eqn:parametricfocusing}) immediately gives
\begin{equation} \label{eqn:queuetail2}
{\cal P}(Q \geq q) \leq K 2^{-(\rho_R - \epsilon) q}.
\end{equation}

\subsection{The high priority stream}

Since the highest priority stream preempts the lower priority stream,
it effectively does not have to share the channel at all. The
queue-length is therefore the same as it would have been for a single
bitstream at rate $R_1$. This establishes the desired result for the
high priority stream.

\subsection{The low priority streams}
Let $Q_H, Q_L$ be the steady-state queue lengths for the high and low
priority queues respectively. Similarly let $D_H, D_L$ be the delays
experienced in the high and low priority queues.
\begin{eqnarray*}
P(D_L \geq d) & = & P(Q_L \geq R_L d) \\
& \leq & P(Q_L + Q_H \geq R_L d) \\
& \leq & K 2^{- (\frac{\alpha(R_H + R_L)}{R_H + R_L} - \epsilon) R_L
  d} \\
& = & K 2^{- (\rho_{HL} - \epsilon) R_L d}
\end{eqnarray*}
Where the final inequality comes from realizing that the combined
queue-length is the same as the queue-length for a single bitstream
arriving with the sum-rate. The last equality comes from plugging in
the definition of $\rho_{HL}$ from (\ref{eqn:rhohldef}) into
(\ref{eqn:queuetail2}). The delay exponent seems to be asymptotically
governed by $\rho_{HL} R_L$. 

The next observation is that the true queue-length exponent must be
monotonically decreasing in rate $R_L$ since increasing the rate of
low-priority message-bit arrivals can only make the low-priority queue
get longer. This allows us to optimize the above bound over all $R_L'
\geq R_L$. Choose $R_L' =  \frac{E_0(\rho)}{\rho} - R_H$ where $\rho
\leq \rho_{HL}$. This ranges $R_L'$ from $R_L$ up to $1 - \beta -
R_H$. The sum rate is $\frac{E_0(\rho)}{\rho}$ and thus $\rho_{HL'} =
\rho$. So the lower-bound on the asymptotic delay error-exponent for
the low-priority bits becomes 
\begin{eqnarray*}
\max_{R_L \leq R_L' \leq 1 - \beta - R_H} \rho_{HL'}R_L'
& = & \max_{\rho \leq \rho_{HL}} \rho(\frac{E_0(\rho)}{\rho} - R_H) \\
& = & \max_{\rho \leq \rho_{HL}} E_0(\rho) - \rho R_H.
\end{eqnarray*}
It is immediately obvious from \cite{gallager} that this can be no
higher than the sphere-packing bound at $R_H$ with equality possible
if the sphere-packing bound at $R_H$ occurs with a $\rho >
\rho_{HL}$. \hfill $\Box$

It turns out that this bound on the low-priority exponent is tight
whenever it hits the sphere-packing bound since the sphere-packing
bound governs the tail of the inter-renewal times for the
high-priority queue. 

\section{Proof of Lemma~\ref{lem:other_stream_error_lemma}}  \label{app:lemma}
Assume all the rates $R_i = R$ for simplicity. First write the
expression corresponding to equation (5) in \cite{ControlPartI} for
the states $i < n$. \cite{ControlPartI} tells us that
$(2+\epsilon_i) = \lambda^{\frac{1}{R_i}}$ and so the virtual
uncontrolled state $\check{X}_t(i) =$
\begin{eqnarray} 
&  & \gamma \lambda^t [\left(\sum_{k=0}^{\lfloor R t \rfloor}
\lambda^{-\frac{k}{R}} S_i(k)\right) \nonumber\\
& + & \left(\sum_{k=0}^{\lfloor R t \rfloor} 
\lambda^{-\frac{k}{R}} p_{1}(\lfloor R t \rfloor - k)
S_{i-1}(k)\right) \nonumber\\
& + & \cdots + 
\left(\sum_{k=0}^{\lfloor R t \rfloor} \lambda^{-\frac{k}{R}}
p_{n-i}(\lfloor R t \rfloor - k) S_{n}(k)\right) \label{eqn:vectorcantorencoding}
\end{eqnarray}
where the $p_k$ represent polynomials that depend on the $A$ matrix.
The key feature of polynomials is that for every $\epsilon$, it is
possible to choose a constant $K_i > 0$ so that $p_k(\tau) \leq K_i 2^{\epsilon
  \tau}$. The maximum possible deviation is bounded by considering the
case in which an error is made on all the bits after a certain point
$t-d$ since the worst case is when every bit that could be wrong is
wrong.

In that worst case, the magnitude of the deviation in $D_j$ due
directly to decoding errors is given by:
\begin{eqnarray*} 
& & \gamma \lambda^t 
\sum_{k=\lceil R (t-d) \rceil}^{\lfloor R t \rfloor}  
\lambda^{-\frac{k}{R}} p_{i-j}(\lfloor R t \rfloor - k) 2 \\
& \leq & 2 K \gamma \lambda^t 
\sum_{k=\lceil R (t-d) \rceil}^{\lfloor R t \rfloor}  
\lambda^{-\frac{k}{R}} 2^{\epsilon (\lfloor R t \rfloor - k)} \\
& \leq & 2 K \gamma 2^{(R\epsilon + \log_2 \lambda)t}
\sum_{k=\lceil R (t-d) \rceil}^{\lfloor R t \rfloor}  
2^{-k (\epsilon + \frac{\log_2 \lambda }{R})} \\ 
& \leq & 
2 K \gamma 2^{(R\epsilon + \log_2 \lambda)t} 
2^{-R(t-d)(\epsilon + \frac{\log_2 \lambda }{R})}
\sum_{k=0}^{\infty} 2^{-k (\epsilon + \frac{\log_2 \lambda }{R})} \\ 
& = & 
K' 2^{(Rt\epsilon + t\log_2 \lambda) - (Rt - Rd)(\epsilon + \frac{\log_2 \lambda }{R})} \\ 
& = & K' 2^{d (\epsilon R + \log_2 \lambda)} \\
& = & K' 2^{d (1 + \frac{\epsilon R}{\log_2 \lambda})\log_2 \lambda }
\end{eqnarray*}
Since $\epsilon$ was arbitrary, choose it so
$\epsilon' = \frac{\epsilon R}{\log_2 \lambda}$. \hfill $\Box$ \vspace{0.15in}

\section{The virtual controlled process} \label{app:virtualcontrol}
{\bf This section is here for the convenience of the reviewers. It is
  a copy of the relevant section from Part I of this paper. It will be
  dropped in the final version of this correspondence.}

The observer that has access to the state and knowledge of the
controls can reconstruct the driving noise $W_t$ since $W_t = X_{t+1}-
\lambda X_t - U_t$. Thus, it has access to the uncontrolled
process \begin{equation}\label{eqn:checkX} 
 \check{X}_{t+1} = \lambda\check{X}_t + W_t.
\end{equation}

The observer acts as though it is working with a virtual controller
through a noiseless channel of finite rate $R$ in the manner. The
resulting bits are sent through the anytime code. The controller
attempts to make the true state behave like the virtual controlled
state by constantly correcting for any erroneous controls that it
might have applied in the past due to tentative bit errors made by the
anytime decoder.

The observer is constructed to keep the state uncertainty at the
virtual controller inside a box of size $\Delta$ by using bits at the
rate $R$. It does this by simulating a virtual process $\bar{X}_t$
governed by:

\begin{equation}\label{eqn:barupdate}
\bar{X}_{t+1} = \lambda \bar{X}_{t} + W_{t} + \bar{U}_t
\end{equation}
where the $\bar{U}_t$ represent the computed actions of the virtual
controller. This gives rise to a virtual undisturbed process
\begin{equation} \label{eqn:XbarU}
X^{\bar{U}}_{t+1} = \lambda X^{\bar{U}}_{t} + \bar{U}_t
\end{equation}
that satisfies the relationship $\bar{X}_t = \check{X}_t +
X^{\bar{U}}_{t}$. The goal is to keep $\bar{X}_t$ within a box
$[-\frac{\Delta}{2},\frac{\Delta}{2}]$, and thereby keep
$-X^{\bar{U}}_{t}$ close to $\check{X}_t$.  

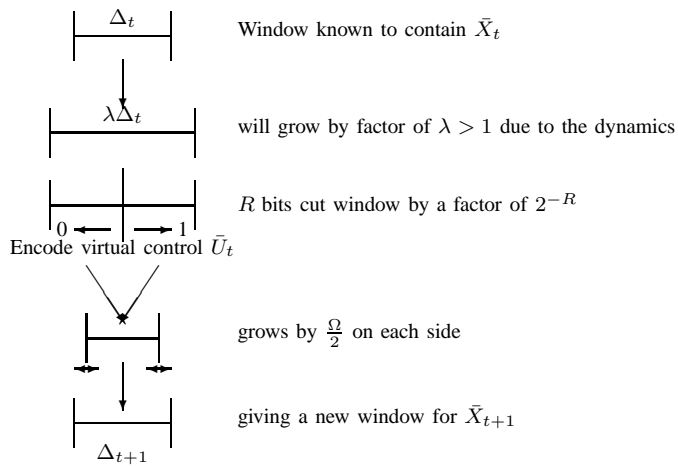
\begin{figure}
\begin{center}
\setlength{\unitlength}{2000sp}%
\begingroup\makeatletter\ifx\SetFigFont\undefined%
\gdef\SetFigFont#1#2#3#4#5{%
  \reset@font\fontsize{#1}{#2pt}%
  \fontfamily{#3}\fontseries{#4}\fontshape{#5}%
  \selectfont}%
\fi\endgroup%
\begin{picture}(6858,5620)(868,-5069)
\put(3226,239){\makebox(0,0)[lb]{\smash{\SetFigFont{8}{14.4}{\rmdefault}{\mddefault}{\updefault}{\color[rgb]{0,0,0}Window known to contain $\bar{X}_t$}%
}}}
\put(3226,-1936){\makebox(0,0)[lb]{\smash{\SetFigFont{8}{14.4}{\rmdefault}{\mddefault}{\updefault}{\color[rgb]{0,0,0}$R$ bits cut window by a factor of $2^{-R}$}%
}}}
\put(1051,-2236){\makebox(0,0)[b]{\smash{\SetFigFont{8}{14.4}{\rmdefault}{\mddefault}{\updefault}{\color[rgb]{0,0,0}0}%
}}}
\put(2551,-2236){\makebox(0,0)[b]{\smash{\SetFigFont{8}{14.4}{\rmdefault}{\mddefault}{\updefault}{\color[rgb]{0,0,0}1}%
}}}
\thinlines
{\color[rgb]{0,0,0}\put(1201,239){\line( 1, 0){1200}}
}%
{\color[rgb]{0,0,0}\put(1201,539){\line( 0,-1){600}}
}%
{\color[rgb]{0,0,0}\put(2401,539){\line( 0,-1){600}}
}%
{\color[rgb]{0,0,0}\put(1801,-61){\vector( 0,-1){600}}
}%
{\color[rgb]{0,0,0}\put(2701,-661){\line( 0,-1){600}}
}%
{\color[rgb]{0,0,0}\put(901,-661){\line( 0,-1){600}}
}%
\thicklines
{\color[rgb]{0,0,0}\put(901,-961){\line( 1, 0){1800}}
}%
\thinlines
{\color[rgb]{0,0,0}\put(1801,-1411){\line( 0,-1){900}}
}%
\thicklines
{\color[rgb]{0,0,0}\put(901,-1861){\line( 1, 0){1800}}
}%
\thinlines
{\color[rgb]{0,0,0}\put(1951,-2161){\vector( 1, 0){450}}
}%
{\color[rgb]{0,0,0}\put(1651,-2161){\vector(-1, 0){450}}
}%
{\color[rgb]{0,0,0}\put(1351,-2611){\vector( 2,-3){484.615}}
}%
{\color[rgb]{0,0,0}\put(2251,-2611){\vector(-2,-3){484.615}}
}%
{\color[rgb]{0,0,0}\put(901,-1561){\line( 0,-1){600}}
}%
{\color[rgb]{0,0,0}\put(2701,-1561){\line( 0,-1){600}}
}%
{\color[rgb]{0,0,0}\put(1351,-3511){\line( 1, 0){900}}
}%
{\color[rgb]{0,0,0}\put(1351,-3211){\line( 0,-1){600}}
}%
{\color[rgb]{0,0,0}\put(2251,-3211){\line( 0,-1){600}}
}%
{\color[rgb]{0,0,0}\put(1801,-3811){\vector( 0,-1){600}}
}%
{\color[rgb]{0,0,0}\put(1201,-3886){\vector(-1, 0){  0}}
\put(1201,-3886){\vector( 1, 0){300}}
}%
{\color[rgb]{0,0,0}\put(2101,-3886){\vector(-1, 0){  0}}
\put(2101,-3886){\vector( 1, 0){300}}
}%
{\color[rgb]{0,0,0}\put(1201,-4561){\line( 1, 0){1200}}
}%
{\color[rgb]{0,0,0}\put(2401,-4261){\line( 0,-1){600}}
}%
{\color[rgb]{0,0,0}\put(1201,-4261){\line( 0,-1){600}}
}%
\put(3226,-3511){\makebox(0,0)[lb]{\smash{\SetFigFont{8}{14.4}{\rmdefault}{\mddefault}{\updefault}{\color[rgb]{0,0,0}grows by $\frac{\Omega}{2}$ on each side}%
}}}
\put(3226,-4561){\makebox(0,0)[lb]{\smash{\SetFigFont{8}{14.4}{\rmdefault}{\mddefault}{\updefault}{\color[rgb]{0,0,0}giving a new window for $\bar{X}_{t+1}$}%
}}}
\put(3226,-961){\makebox(0,0)[lb]{\smash{\SetFigFont{8}{14.4}{\rmdefault}{\mddefault}{\updefault}{\color[rgb]{0,0,0}will grow by factor of $\lambda>1$ due to the dynamics }%
}}}
\put(1801,-5011){\makebox(0,0)[b]{\smash{\SetFigFont{8}{14.4}{\rmdefault}{\mddefault}{\updefault}{\color[rgb]{0,0,0}$\Delta_{t+1}$}%
}}}
\put(1801,-2461){\makebox(0,0)[b]{\smash{\SetFigFont{8}{14.4}{\rmdefault}{\mddefault}{\updefault}{\color[rgb]{0,0,0}Encode virtual control $\bar{U}_t$}%
}}}
\put(1801,-811){\makebox(0,0)[b]{\smash{\SetFigFont{8}{14.4}{\rmdefault}{\mddefault}{\updefault}{\color[rgb]{0,0,0}$\lambda\Delta_t$}%
}}}
\put(1801,389){\makebox(0,0)[b]{\smash{\SetFigFont{8}{14.4}{\rmdefault}{\mddefault}{\updefault}{\color[rgb]{0,0,0}$\Delta_t$}%
}}}
\end{picture}
\end{center}
\caption{Virtual controller for R=1. How the virtual state
$\bar{X}$ evolves.}
\label{fig:causalmarkovcode}
\end{figure}
 
Because of the rate constraint, the virtual control $\bar{U}_t$ takes
on one of $2^{\lfloor R(t+1) \rfloor - \lfloor Rt \rfloor}$
values. For simplicity of exposition, ignore the integer effects
and consider it to be one of $2^R$ values and proceed by
induction. Assume that $\bar{X}_t$ is known to lie within
$[-\frac{\Delta}{2},\frac{\Delta}{2}]$. Then $\lambda\bar{X}_t$ will
lie within $[-\frac{\lambda \Delta}{2},\frac{\lambda \Delta}{2}]$. By
choosing $2^R$ control values uniformly spaced within that interval,
it is guaranteed that $\lambda \bar{X}_t + \bar{U}_t$ will lie within
$[-\frac{\lambda \Delta}{2^{R+1}},\frac{\lambda
\Delta}{2^{R+1}}]$. Finally, the state will be disturbed by $W_t$ and
so $\bar{X}_{t+1}$ will be known to lie within $[-\frac{\lambda
\Delta}{2^{R+1}} - \frac{\Omega}{2}, \frac{\lambda \Delta}{2^{R+1}} +
\frac{\Omega}{2}]$.

Since the initial condition has no uncertainty, induction will be
complete if
\begin{equation} \label{eqn:inductioncondition}
\frac{\lambda}{2^{R}} \Delta + \Omega \leq \Delta
\end{equation}
To get the minimum $\Delta$ required as a function of $R$, we can
solve for (\ref{eqn:inductioncondition}) being an equality. This
occurs when $\Delta = \frac{\Omega}{1 - \lambda 2^{-R}}$ for every
case where $R > \log_2 \lambda$. Since the slope $\frac{\lambda}{2^R}$
on the left hand side of (\ref{eqn:inductioncondition}) is less than
$1$, any larger $\Delta$ also works.

\section*{Acknowledgments}
The authors would like to thank Mukul Agarwal for comments on earlier
versions of this correspondence. We thank Nicola Elia for several
constructive discussions about the subject matter and thank Sekhar
Tatikonda for many discussions over a long period of time which have
influenced this work in important ways. We also thank the anonymous
reviewer and associate editor for comments that improved the clarity
and presentation of these results.

\bibliographystyle{IEEEtran}
\bibliography{IEEEabrv,./MyMainBibliography}

\end{document}